# Evidence that the AGN dominates the radio emission in $z \sim 1$ radio-quiet quasars

Sarah V. White[1,2]*, Matt J. Jarvis[2,3], Eleni Kalfountzou[4,5], Martin J. Hardcastle[5], Aprajita Verma[2], José M. Cao Orjales[5], Jason Stevens[5]

[1] *International Centre for Radio Astronomy Research (ICRAR), Curtin University, Bentley, WA 6102, Australia*
[2] *Astrophysics, University of Oxford, Denys Wilkinson Building, Keble Road, Oxford, OX1 3RH, UK*
[3] *Department of Physics, University of the Western Cape, Bellville 7535, South Africa*
[4] *European Space Astronomy Centre (ESAC), Villanueva de la Cañada, E-28692 Madrid, Spain*
[5] *Centre for Astrophysics Research, School of Physics, Astronomy and Mathematics, University of Hertfordshire, Hatfield, Herts, AL10 9AB, UK*

**ABSTRACT**
In order to understand the role of radio-quiet quasars (RQQs) in galaxy evolution, we must determine the relative levels of accretion and star-formation activity within these objects. Previous work at low radio flux-densities has shown that accretion makes a significant contribution to the total radio emission, in contrast with other quasar studies that suggest star formation dominates. To investigate, we use 70 RQQs from the *Spitzer-Herschel* Active Galaxy Survey. These quasars are all at $z \sim 1$, thereby minimising evolutionary effects, and have been selected to span a factor of $\sim 100$ in optical luminosity, so that the luminosity dependence of their properties can be studied. We have imaged the sample using the Karl G. Jansky Very Large Array (JVLA), whose high sensitivity results in 35 RQQs being detected above $2\sigma$. This radio dataset is combined with far-infrared luminosities derived from grey-body fitting to *Herschel* photometry. By exploiting the far-infrared–radio correlation observed for star-forming galaxies, and comparing two independent estimates of the star-formation rate, we show that star formation alone is not sufficient to explain the total radio emission. Considering RQQs above a 2-$\sigma$ detection level in both the radio and the far-infrared, 92 per cent are accretion-dominated, and the accretion process accounts for 80 per cent of the radio luminosity when summed across the objects. The radio emission connected with accretion appears to be correlated with the optical luminosity of the RQQ, whilst a weaker luminosity-dependence is evident for the radio emission connected with star formation.

## 1 INTRODUCTION

Two crucial aspects of galaxy evolution are black-hole accretion and star formation. That the two processes are connected is suggested by the fact that the cosmic history of star formation (e.g. Madau et al. 1996; Hopkins & Beacom 2006) follows a similar trajectory to that of the cosmic history of accretion (e.g. Ueda et al. 2003; Wolf et al. 2003). They both peak in activity at $z \sim 1$–2 (Madau & Dickinson 2014), and the interaction of these processes is thought to explain, for example, why black-hole mass and the velocity dispersion of the bulge are well correlated for a wide range of galaxy types (e.g. Gültekin et al. 2009). However, it is not yet well understood how the two phenomena interact or whether their similar histories are simply due to a common factor: the availability of gas.

The interaction of accretion and star formation is mediated through 'feedback' mechanisms. For example, the radiative output of an active galactic nucleus (AGN) may heat nearby gas, thereby suppressing star formation. As such, this process is a form of 'negative' feedback, as it limits further growth of the system. Alternatively, 'positive' feedback may arise, where mechanisms connected to accretion or star formation *promote* additional growth. In the case of an AGN, its jet could de-stabilise a molecular cloud, causing it to collapse and form stars. The impact of such feedback is clearly dependent on the environment, with triggered star-formation occurring on nuclear (Davies et al. 2006), galactic (Ishibashi & Fabian 2012), and intracluster (Croft et al. 2006) scales.

Other sources of feedback include supernovae, which expel material that might otherwise accrete onto the central black-hole or be used to fuel star formation. This is another example of negative feedback, as it precludes further growth of the black hole or the host galaxy. The remnants left behind by supernovae in star-forming regions accelerate relativistic electrons, resulting in synchrotron radiation that is visible at radio frequencies. Therefore, since accretion also produces synchrotron emission, radio observations can provide an unobscured view of both AGN and star-forming populations, unaffected by dust.

* sarah.white@icrar.org

As we probe fainter radio flux-densities, studies of radio sources suggest that the dominant population switches from radio-loud AGN to star-forming galaxies (Hopkins et al. 2003; Wilman et al. 2008; Condon et al. 2012). However, there may still be ongoing star-formation in the host galaxies of AGN (e.g. Canalizo & Stockton 2001; Netzer et al. 2007; Silverman et al. 2009). Kimball et al. (2011) and Condon et al. (2013) investigated this further by studying the radio emission from samples of optically-selected quasars. For those with faint radio emission, they proposed that this emission originates from star formation in the host galaxy, rather than the AGN. In contrast, a recent analysis by White et al. (2015) indicates that black-hole accretion makes a significant contribution to the total radio emission in radio-quiet quasars (RQQs). A similar approach is taken by Zakamska et al. (2016), who use different SFR indicators to demonstrate that star formation is insufficient to explain the total radio emission. In addition, Herrera Ruiz et al. (2016) find evidence that accretion is dominating the radio emission in RQQs, by using Very Long Baseline Interferometry (VLBI) to directly measure the radio luminosity associated with their compact cores. We will further investigate the origin of the radio emission in RQQs in the present work, using a sample at $z \sim 1$.

Although the above studies emphasise the contribution of the AGN to the radio emission, they do not preclude quasar host-galaxies exhibiting star formation. The latter can be traced at other wavelengths, such as in the far-infrared (FIR). When considering the origin of the FIR continuum in quasars, Sanders et al. (1989) concluded that the emission over 2–1000 μm was mainly due to re-radiation from a warped disc, heated by the AGN. Today the general consensus agrees with Rowan-Robinson (1995) and Haas et al. (1999), who argued that it is star formation in the quasar hosts that produces the bulk of the FIR emission in quasars. However, the relative contributions to the emission by star formation and the dusty torus surrounding the AGN (e.g. Mullaney et al. 2011) may also depend on redshift.

Several studies, such as that by Rosario et al. (2013), use FIR measurements to quantify the level of star formation in quasars. They show that quasar hosts and typical, massive, star-forming galaxies have mean SFRs that are consistent with one another. In addition, Bonfield et al. (2011) use data from the *Herschel Space Observatory* (Pilbratt et al. 2010) and find that there is an appreciable correlation between the star-formation luminosity (traced by FIR emission) and the accretion luminosity (traced by optical emission). However, the samples in these studies suffer from Malmquist bias, intrinsic to flux-density limited samples.

### 1.1 Disentangling the radio emission due to accretion and star formation

In order to study how star formation in galaxies correlates with the amount of accretion activity, contributions from the two different processes to the total radio emission must first be disentangled. This is particularly difficult in objects without obvious jets, as morphology cannot then be used to distinguish between star-formation and accretion components. Therefore, for an accurate investigation of the origin of the radio emission, a multi-wavelength dataset is required.

In the optical, the 'Baldwin, Phillips & Terlevich' (BPT) diagram is traditionally used to provide object classification via diagnostic line-ratios (Baldwin et al. 1981; Kewley et al. 2001), allowing AGN to be separated from starbursts. The emission lines used for this method are sufficiently close that the effect of (moderate) dust obscuration is negated. However, the presence of dust means that ultra-violet and optically-selected samples are prone to missing the most-obscured objects. In such objects, without the necessary emission lines in the spectrum, the BPT diagram is insufficient to separate emission of AGN origin from that due to star formation. Furthermore, when the nuclear emission is dominant (as is the case for quasars), the bright continuum and broad emission lines can greatly outshine the emission from the host galaxy, even if it is strongly star-forming. Meanwhile, mid-infrared equivalents of the BPT diagram and spectral decomposition have also been used to disentangle AGN-related emission from star formation, using *Infrared Space Observatory* and *Spitzer* spectroscopy (e.g. Laurent et al. 2000; Sturm et al. 2002; Verma et al. 2005, and references therein; Veilleux et al. 2009; Petric et al. 2011; Hernán-Caballero et al. 2015). However, we cannot use such a method for the sample of RQQs studied in this paper, as they are faint in the mid-infrared, and beyond the sensitivity of current mid-infrared spectroscopic instruments (i.e. SOFIA/FORCAST).

Instead, we will use FIR data from *Herschel* in combination with deep radio data to break the degeneracy between the accretion and star-formation contributions. This is possible because, as shown by Helou et al. (1985) and de Jong et al. (1985), a tight correlation exists between the FIR and the radio, known as the far-infrared–radio correlation (FIRC). This enables us to determine how much of the total radio emission is due to star formation, with any 'excess' radio emission then being attributed to the AGN.

The reason for this correlation is that both types of emission are produced by star formation (Bell 2003) – a process that heats the large quantities of dust present in star-forming regions. Being at cool temperatures ($\sim 30$ K), this dust then radiates in the FIR via black-body radiation. As for the radio emission, supernova remnants are naturally co-located with regions of star formation. Electrons are accelerated rapidly as they traverse these shock fronts, thus producing synchrotron radiation (Lisenfeld et al. 1996; Lacki et al. 2010). However, one might expect the correlation to be affected by the delay between stars forming and supernovae occurring. This is not the case, though, as the supernovae in question are only produced by stars more massive than $\sim 8$ M$_\odot$. These stars have lifetimes of $\leq 3 \times 10^7$ yr, which is a factor ten shorter than the lifetimes of relativistic electrons (Condon 1992). Therefore, the radio emission gives an integrated view of the star formation history of a galaxy, between the time of observation and $\sim 0.3$ Gyr previously.

### 1.2 Paper outline

In this paper we present an investigation of the source of the radio emission in radio-quiet quasars (RQQs), which complements the study (carried out over a larger redshift range) by White et al. (2015). By using a sample that is restricted to $z \sim 1$, we are able to minimise the impact of any evolutionary effects on our final results. Our sample has *Herschel* photometry for the RQQs, which span a wide range in optical luminosity. Thus the level of radio emission associated with star formation can be studied as a function of absolute *i*-band magnitude. Furthermore, the separation of star for-mation and accretion, in terms of their radio emission, allows the connection between the two processes to be tested.

Section 2 outlines how the sample is selected, and Section 3 describes the multi-wavelength data that we use. This includes new radio observations from the Karl G. Jansky Very Large Array (JVLA), whose reduction is detailed in Section 4. Analysis of the radio emission is presented in Section 5, and that for the far-infrared emission is described in Section 6. Section 7 explains how the radio and far-infrared results are used together to investigate the level of star formation in the quasars. The trend in star-formation level with optical luminosity is presented in Section 8, along with that for the accretion-connected radio luminosity. This is followed by discussion in Section 9, and our conclusions in Section 10. AB magnitudes are used throughout this paper, and we use a $\Lambda$CDM cosmology, with $H_0 = 70 \, \mathrm{km \, s^{-1} \, Mpc^{-1}}$, $\Omega_m = 0.3$, $\Omega_\Lambda = 0.7$.

## 2 SAMPLE SELECTION

Investigations of quasar properties that use samples covering a wide redshift range are subject to $K$-correction uncertainties and evolutionary effects. These are exacerbated by Malmquist bias. Therefore, careful study of objects belonging to a single epoch is an important step towards understanding the radio emission from the quasar population. These findings, including the way in which properties vary as function of optical luminosity, can then be incorporated into how the underlying physical processes evolve over cosmic time.

A sample of 70 RQQs is used for the work presented in this paper. These are taken from the *Spitzer-Herschel* Active Galaxy Survey (Kalfountzou et al., in prep.), which consists of quasars that are selected from the Sloan Digital Sky Survey (SDSS) using multi-colour criteria. The quasars are restricted to a thin redshift slice of $0.9 < z < 1.1$, so chosen because: (a) this is the minimum redshift at which there are sufficient high-luminosity quasars for comparison to bright objects at higher redshifts, and (b) the targets are still close (i.e. bright) enough that a large sample of them can be studied in a reasonable amount of time.

The quasars span nearly 5 magnitudes in optical luminosity ($1.1 \times 10^{11} < L_{\mathrm{opt}}/\mathrm{L}_\odot < 9.6 \times 10^{12}$), and the FIRST (Faint Images of the Radio Sky at Twenty-cm; White et al. 1997), NVSS (NRAO VLA Sky Survey; Condon et al. 1998), and WENSS (Westerbork Northern Sky Survey; Rengelink et al. 1997) radio catalogues were used to identify those that were classed as radio-loud. Quasars undetected in these radio surveys are therefore classed as radio-quiet. These were then divided into four bins in bolometric luminosity, and a sample randomly selected such that there are around 20 RQQs per bin. As these sources are undetected in FIRST, which has a flux-density limit of 1 mJy beam$^{-1}$, they were targeted using the JVLA (Section 3.3).

## 3 DATA

As illustrated by Best & Heckman (2012), Stern et al. (2014) and White et al. (2015), ancillary data are crucial to large-sample AGN studies; observations at different wavelengths offer different insights into AGN physics, therefore allowing various galaxy-evolution processes to be disentangled from one another. Whilst mid-infrared data and X-ray data exist for these quasars, in the form of *Spitzer* and *XMM-Newton* observations, we describe below only the datasets that we used in the current investigation. Kalfountzou et al. (in prep.) will present a comprehensive analysis of the sample properties across the different wavebands.

### 3.1 Optical: SDSS

Optical photometry (*ugriz*) from the SDSS is used for the pre-selection of the quasars (Richards et al. 2002; Schneider et al. 2005). Follow-up spectroscopy from the fifth data release (DR5; Adelman-McCarthy et al. 2007) provides accurate redshifts, and allows absolute magnitudes to be calculated. We use the *i* band here, as this is dominated by thermal emission from the accretion disc and is less susceptible to dust than bluer bands. We also use black-hole masses for this sample, calculated by Falder et al. (2010), based on the virial-mass estimate technique of McLure & Jarvis (2002), with Mg$\textsc{ii}_{2799}$ broad lines from SDSS DR5 spectroscopy.

### 3.2 Far-infrared: *Herschel* observations

To estimate the level of star formation in the quasars, we use 5 FIR photometric bands from *Herschel*. These are at 70 and 160 μm using the Photodetector Array Camera and Spectrometer (PACS) instrument (Poglitsch et al. 2010), and at 250, 350 and 500 μm using the Spectral and Photometric Imaging REceiver (SPIRE) instrument (Griffin et al. 2010). Some of the RQQs in the sample lie in fields with existing coverage from *Herschel*, including fields of the *Herschel* Multi-tiered Extragalactic Survey (Oliver et al. 2012), and the remaining measurements (for 50 of the 70 sources) were obtained as part of a targeted program (PI: Stevens). The mean 1-$\sigma$ photometric uncertainties for the PACS and SPIRE bands are 2.56, 5.27, 6.29, 5.38, 7.09 mJy, estimated via the method of Elbaz et al. (2011) and Pascale et al. (2011), and the beam sizes range from 5 to 36 arcsec at full-width half-maximum. In the 250 μm band, 69 per cent of the quasars are detected above 2$\sigma$, and 49 per cent are detected above 3$\sigma$. This is a higher detection rate than that found by Pitchford et al. (2016), largely due to our *Herschel* data being a factor ~ 2 deeper, and their sample being dominated by quasars at $z > 1$. In addition, the detection rate may be influenced by our quasars having an artificially uniform distribution in optical luminosity, as a result of how they have been selected (Section 2). All maps were reduced using pipelines within the *Herschel* Interactive Processing Environment (HIPE; Ott 2010), as detailed by Kalfountzou et al. (submitted).

### 3.3 Radio: JVLA observations

We targeted all RQQs with the JVLA. An observation time of 25 minutes per quasar was chosen in order to achieve an rms noise level of 12.5 μJy beam$^{-1}$. In addition to this, 10 minutes per source for overheads and calibration were required, leading to 42 hours of observations in total (VLA-12B-115, PI: Jarvis). These were taken between 4th November 2012 and 9th December 2012. Visibilities were recorded every second for 16 spectral windows, each having 64 channels of 1-MHz bandwidth. The most-extended configuration of the JVLA, the 'A' array, was used to obtain the highest resolution. For the L band, this results in a synthesised beam with a Half-Power Beam Width (HPBW) of 1.3 arcsec (similar to the size of a typical galaxy at $z \sim 1$). Note that, in the present work, reference to the radio flux-density (or luminosity) at 1.5 GHz corresponds to the new L-band measurement, which covers 1–2 GHz.

## 4 JVLA DATA REDUCTION

Our JVLA program consisted of 11 scheduling blocks, each (typically) containing data for one flux calibrator, three phase calibrators, and seven science targets. The reduction of these blocks was carried out using the Common Astronomy Software Applications (CASA) package (McMullin et al. 2007). This section provides further details of the processing involved.

### 4.1 Calibrating JVLA data

Before running a particular observation block through the JVLA calibration pipeline (Version 1.2.0 for use with CASA 4.1.0), the data were Hanning-smoothed and flagged for initial radio-frequency interference (RFI). The pipeline performs the standard calibration steps, and for 10 of the 11 scheduling blocks, this included using the calibrator J1331+3030 (3C 286) for both flux and bandpass calibration. In the case of the remaining block, J0137+331 (3C 48) was used.

The diagnostic plots output by the pipeline indicated that additional flagging was required to eliminate corrupted visibilities. Furthermore, a number of phase calibrators showed unexpected variation in amplitude and phase with $u$-$v$ distance. This helped to pinpoint faulty antennae, but also highlighted calibrators whose large-scale structure was starting to be resolved. The 'uvmin' and 'uvmax' values within the VLA Calibrator Manual[1] (in addition to the pipeline's output) helped to inform the appropriate $u$-$v$ range that should be used for re-calibration.

We emphasise that these $u$-$v$ restrictions were applied to the phase calibrator *only*. Since gain-calibration solutions are antenna-based rather than baseline-based, such flagging does not affect which baselines are present once the solutions are applied to the target data. If this was not the case, then the sensitivity to extended structure would be lost and a radio jet (if present) would not be imaged. So, with $u$-$v$ ranges specified for the phase calibrators, the pipeline was run a second time to re-calibrate the visibilities.

### 4.2 Imaging JVLA data

Imaging of the targets was carried out using CASA's standard CLEAN algorithm, for which we specified a pixel size of 0.25 arcsec. This enabled good sampling of the synthesised beam, with just over 5 pixels across the full-width half-maximum. The setting of initial CLEAN boxes was informed by radio cutouts from FIRST, and additional boxes were added in the interactive mode as fainter emission became apparent. We used a 'robust' parameter of 0.5 for Briggs weighting of the visibilities, as this gives the best compromise between sensitivity and angular resolution. As a result of the JVLA being upgraded, the multi-frequency synthesis CLEAN mode needs to be implemented, which mitigates the effect of bandwidth smearing on the visibilities.

Due to processing restrictions, preliminary images were limited to $1000 \times 1000$ pixel in size. These were of sufficient quality in most cases, but at 4.2 arcmin across, this meant that only a fraction of the primary-beam size (30 arcmin at L band) is covered by each image. Therefore, any bright objects that are external to this field-of-view, either present in the primary beam or the sidelobes of the beam response function, may make a non-negligible contribution to the flux density measured for the target at the pointing centre (Condon et al. 1998). To alleviate this issue, such objects were imaged so that CLEAN boxes could be placed around them, allowing their brightness to be correctly modelled via CLEAN components.

Taking into account the signal-to-noise ratio (SNR) of the target, further processing was carried out for images that displayed striping or radial artefacts. The first step was to generate images with a field-of-view covering $75 \times 75$ arcmin$^2$, this being 2.5 times the primary beam. For such wide-field images, it is necessary to employ a $w$-projection algorithm during the CLEAN. A lower spatial resolution of 1.5 arcsec per pixel was deemed adequate since these images were simply to give an idea of what emission lay outside the central $250 \times 250$ arcsec$^2$ region. Furthermore, this imaging was run non-interactively (i.e. no CLEAN boxes were set), since the reduction in processing time far-outweighed the recovery of any faint emission, which would likely make a negligible contribution to the flux density at the pointing centre.

Strong sources in the primary beam and sidelobes were identified, and their position and extent in emission used to define an 'outlier file' for each target. This allows outliers to be CLEANed and imaged in conjunction with the main $1000 \times 1000$ pixel image. However, in some cases it was clear that the artefacts originated from a source just beyond the $250 \times 250$ arcsec$^2$ covered by the original image. For these targets the main imaging-region was extended slightly to include the outlier emission.

A substantial improvement in imaging quality is seen when outlier files are used for sources detected outside the central region. The resulting smoother fields also lend greater credibility to cases where the pointing-centre flux-density is at the level of the noise.

### 4.3 Final radio measurements

For each image we measure the peak flux-density of the target and record the noise level over a manually-defined region, avoiding all sources. These values are provided in Table A1, where we use the measured rms in the calculation of the SNR. Due to the reduction in sensitivity caused by flagged data, bright objects, and direction-dependent effects (resulting in residual phase and amplitude errors), it is expected that the final images should have a noise level that is 2–3 times the theoretical rms value. In the resulting images, 35 RQQs have a $\geq 2$-$\sigma$ detection (of which 30 are detected at $\geq 3\sigma$), and zoomed-in images for these objects are presented in Figs. 1–3. The five RQQs that have a level of detection between 2 and $3\sigma$ are: SDSS082229.78+442705.2, SDSS092753.52+053637.0, SDSS093023.28+403111.0, SDSS102349.40+522151.2, and SDSS104659.37+573055.6. Although these are borderline 'detections', we proceed with $2\sigma$ as the detection threshold to avoid overcomplicating the analyses (specifically those in Section 7 and Section 8).

## 5 ANALYSIS OF THE RADIO EMISSION

Using the new JVLA images (Section 4.3), the radio flux-density distribution of the RQQs is presented in the following subsection, followed by investigation of how the radio luminosity correlates with optical luminosity.

### 5.1 Radio flux-density measurement

To investigate the radio flux-density distribution of the RQQs, we use the single-pixel flux-densities extracted from the JVLA images at the pointing centre. For detected objects, this value is superseded by the flux density extracted where the target's radio emission *peaks* (Table A1). A tolerance of 1 arcsec from the pointing centre is used to account for any pointing error in the observations, although such an offset is seen for only one of the detected targets, and is more likely due to a phase-calibration error. In each case, the radio emission from well-detected objects appears as an unresolved point-source, with none of the RQQs showing any extended emission. Resolution better than the present 1.3 arcsec would be needed to distinguish between unresolved emission from the central engine and any small radio-jets. We also create a 'random' sample of radio flux-densities by extracting 1000 single-pixel values from each JVLA image at random positions. This is done so that the different noise properties of each of the radio images can be taken into consideration. To ensure that the extracted flux-densities are not correlated with that of the source in question (due to being covered by the same synthesised beam), these pixels are constrained to lie between 3 and 9 arcsec from the pointing centre.

---

[1] http://www.vla.nrao.edu/astro/calib/manual/index.shtml

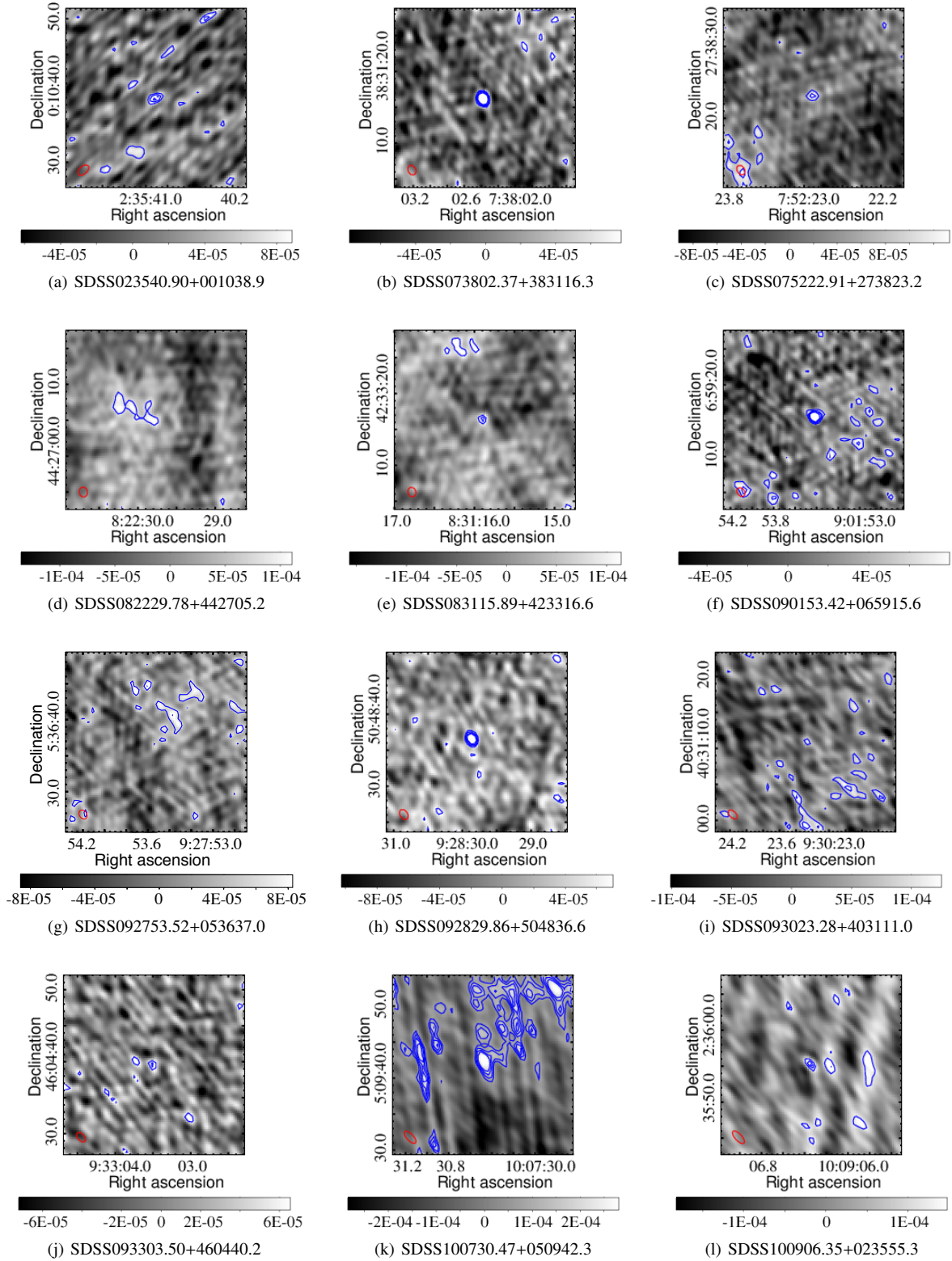

**Figure 1.** JVLA imaging of the RQQs, with each target at the pointing centre. Each of the objects in this figure are deemed 'detections', at the 2-$\sigma$ level. The beamsize is illustrated by the red ellipse in the bottom left-hand corner of each image, and the greyscale bar represents the flux density in units of Jy beam$^{-1}$. The blue contours are at the levels of 2, 3, 4, and 5 $\sigma$, calculated using the rms value.

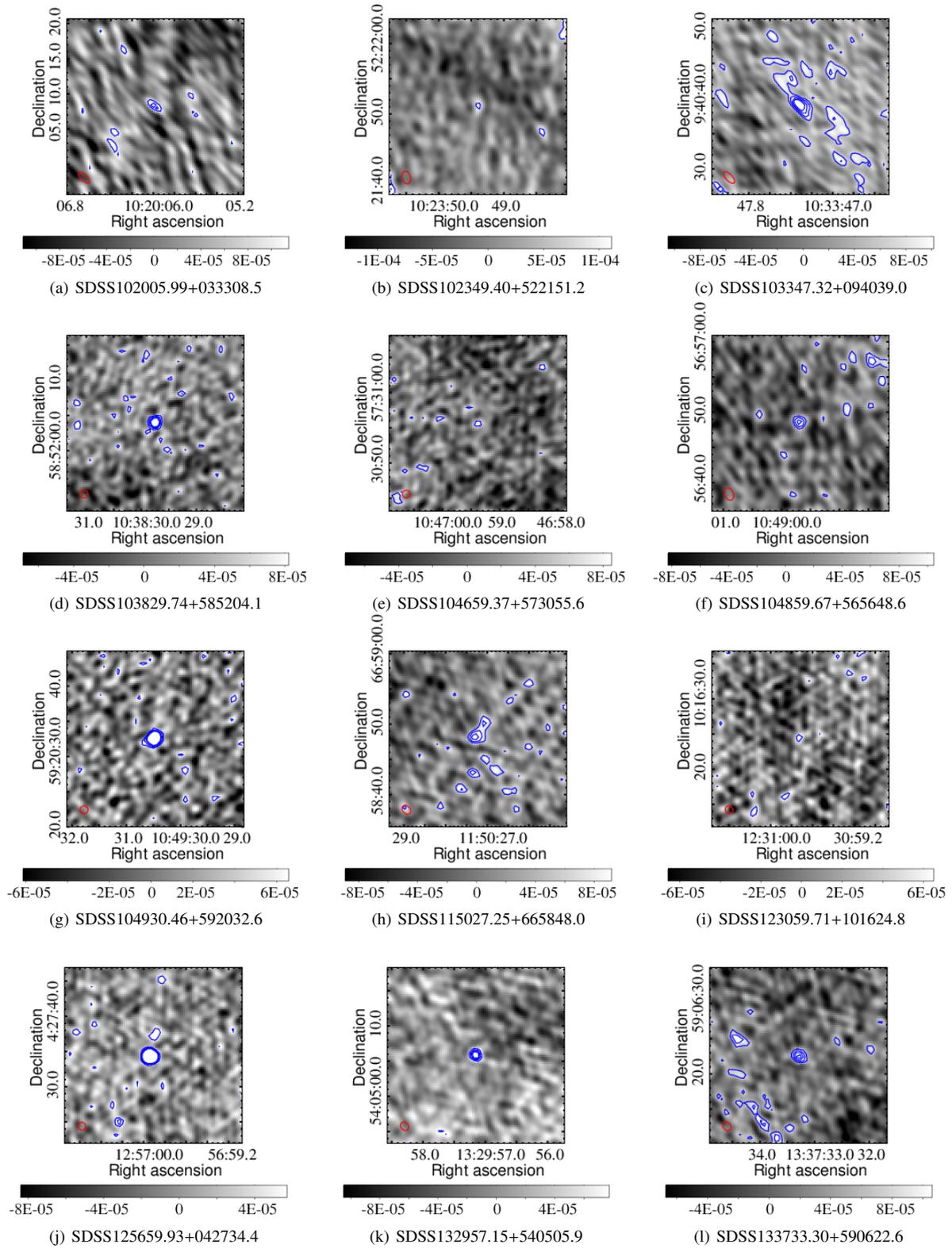

**Figure 2.** JVLA imaging of the RQQs, with each target at the pointing centre. Each of the objects in this figure are deemed 'detections', at the 2-$\sigma$ level. The beamsize is illustrated by the red ellipse in the bottom left-hand corner of each image, and the greyscale bar represents the flux density in units of Jy beam$^{-1}$. The blue contours are at the levels of 2, 3, 4, and 5 $\sigma$, calculated using the rms value.

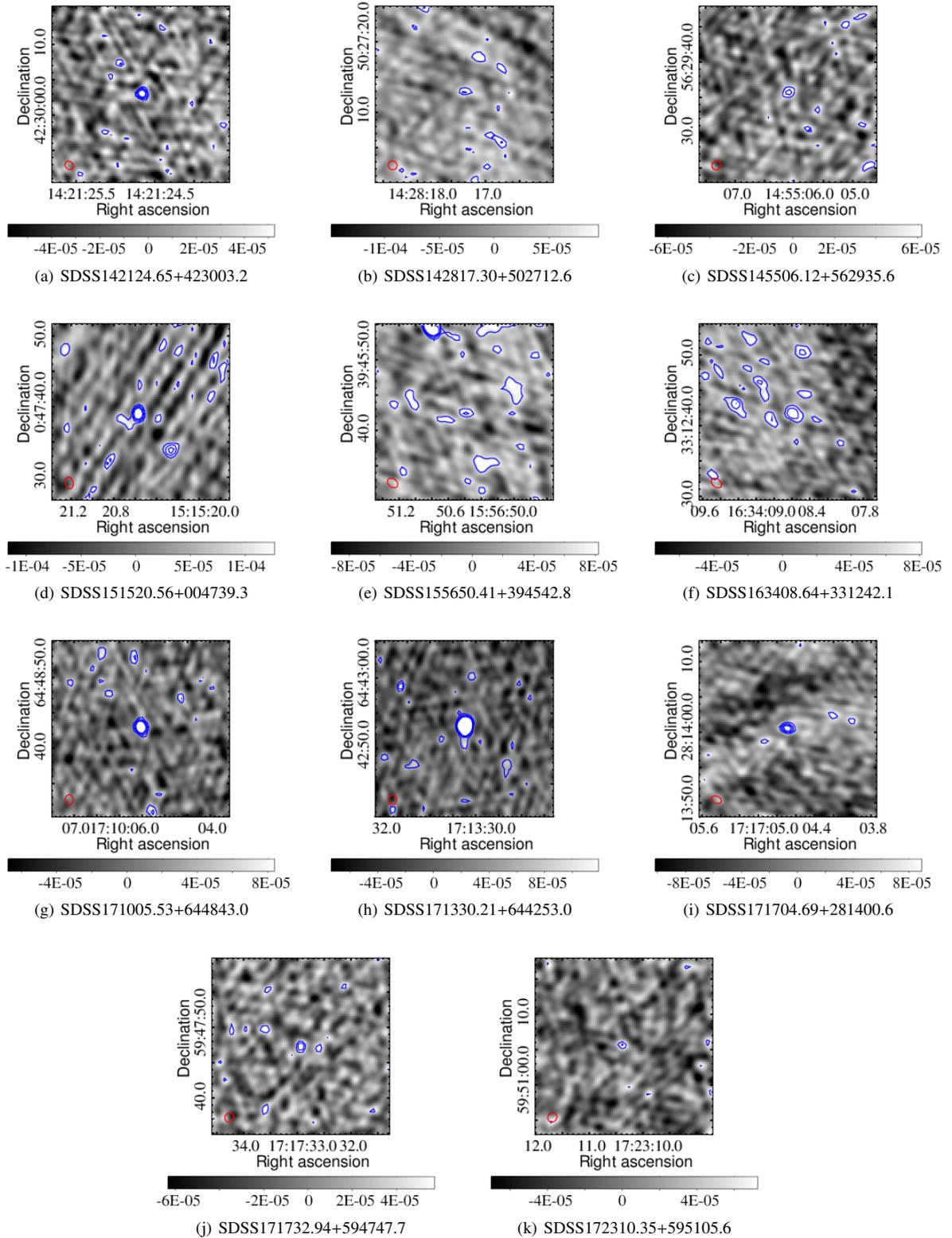

**Figure 3.** JVLA imaging of the RQQs, with each target at the pointing centre. Each of the objects in this figure are deemed 'detections', at the 2-$\sigma$ level. The beamsize is illustrated by the red ellipse in the bottom left-hand corner of each image, and the greyscale bar represents the flux density in units of Jy beam$^{-1}$. The blue contours are at the levels of 2, 3, 4, and 5 $\sigma$, calculated using the rms value.

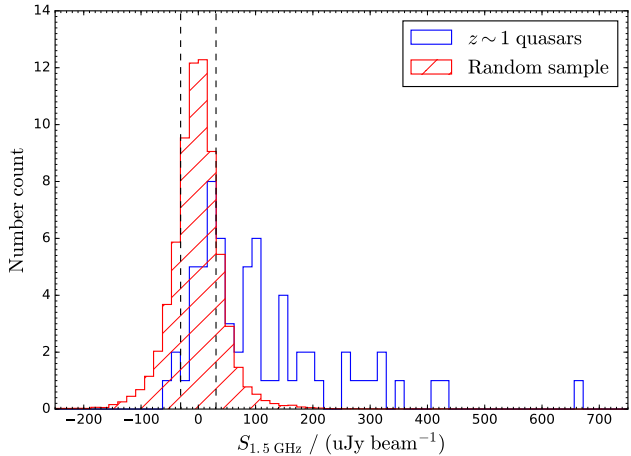
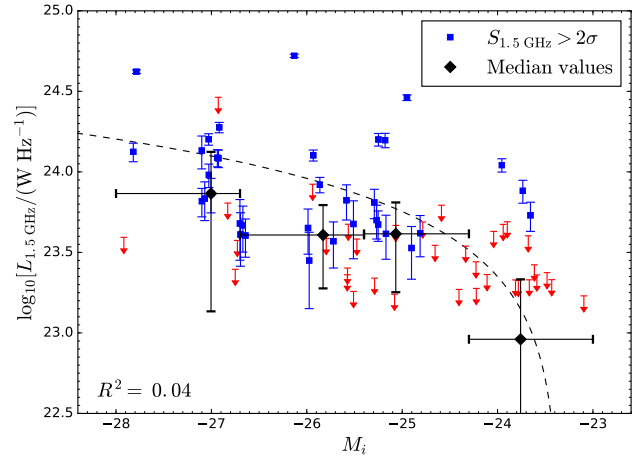

**Figure 4.** The radio flux-density distribution for the RQQs (blue histogram). The red (hatched) histogram corresponds to the random sample, where the flux densities are extracted from pixels 3 to 9 arcsec away from the quasar positions. (Their number counts have been scaled to aid comparison.) The black, dashed lines demarcate the median rms noise level ($\pm 30.9\,\mu$Jy beam$^{-1}$), found by averaging over all 70 JVLA images. Note that three quasars lie beyond the radio flux-density range shown, two of which are brighter than 750 $\mu$Jy beam$^{-1}$.

**Figure 5.** A linear regression analysis, performed between the radio luminosity ($L_{1.5\,\text{GHz}}$) and the absolute $i$-band magnitude ($M_i$) of the RQQs, indicates a line of best-fit given by: $L_{1.5\,\text{GHz}} = (-3.44 \pm 0.87) \times 10^{23} M_i - (8.03 \pm 2.20) \times 10^{24}$. This is the dashed line in this figure, following conversion into log–linear space. The fitting incorporates uncertainties in $L_{1.5\,\text{GHz}}$, and the associated coefficient of determination ($R^2$) is given in the bottom left-hand corner. Blue squares correspond to objects detected in the radio above 2$\sigma$, and red arrows represent 2-$\sigma$ upper limits for objects below this detection threshold. Overplotted are the median luminosities using all objects (black diamonds), derived from the measured flux-densities, binned in $M_i$. They are offset from the dashed line because the latter is determined via mean luminosities. The horizontal error-bars indicate the ranges of the $M_i$ bins (Table 1), and uncertainties on the median radio-luminosities are given by the ordinate error-bars. Note that the values of $L_{1.5\,\text{GHz}}$, even if negative, are used for the linear regression analysis and the calculation of the median luminosities, rather than the upper-limit values.

The 35 quasars detected at the $\geq 2$-$\sigma$ level give rise to the long, positive tail seen in the flux-density distribution (Fig. 4). The noise levels of the individual images, for each of the 70 quasars in the full sample, are averaged to calculate a median rms noise level of 30.9 $\mu$Jy beam$^{-1}$. In Table 1 we provide average flux-densities, for the sample as a whole and when the quasars are binned according to their absolute $i$-band magnitude. Considering the full sample, the mean flux-density is $123.3 \pm 27.6\,\mu$Jy beam$^{-1}$. This is within the 1-$\sigma$ error of the finding by Falder et al. (2010), who used the same sample of RQQs to study the relationship between AGN activity and environmental density at $z \sim 1$. By performing a stacking analysis, they found the average flux-density at 1.4 GHz to be $0.10 \pm 0.02$ mJy.

### 5.2 Radio properties as a function of optical luminosity

Next we determine whether there is a correlation between the radio luminosity and the absolute $i$-band magnitude, which would be expected for RQQs if they exhibit the same accretion processes as radio-loud quasars (e.g. Serjeant et al. 1998; Punsly & Zhang 2011). This is because the amount of radio emission due to accretion may be coupled to the optical luminosity of the accretion disc, due to thermal emission that is produced as material is transported inwards. However, this correlation may still be weak, as the radio emission is thought to be dependent on the accretion of *magnetic flux*, rather than the simple accretion of material (Sikora & Begelman 2013). Furthermore, the radio luminosity from radio lobes is complicated by a dependence on both evolutionary and environmental effects (Hardcastle et al. 2013). For example, Shabala & Godfrey (2013) investigate the connection between radio luminosity and the kinetic power of radio jets, taking into account the source age and the degree of confinement. This could be extended to RQQs, the idea being that the source is young and a jet has not 'turned on' yet.

Alternatively, a different mechanism may exist for RQQs, as suggested by the stacking analysis by Fernandes et al. (2011) on optically-selected quasars, which include the RQQs presented here. Their result lies well above the minimum accretion-rate envelope defined by the radio-loud sources, but this stacked detection could be skewed by a few bright quasars. (The radio data they used came from FIRST, which is not deep enough to study the distribution on an individual basis.)

**Table 1.** Stacked radio flux-densities for the quasars, as a whole sample and when binned in absolute *i*-band magnitude, $M_i$. The error in the median flux-density is estimated via the bootstrap method, using 1000 re-samples of the data.

| Sample | Number of objects | Median flux-density (μJy beam$^{-1}$) | Mean flux-density (μJy beam$^{-1}$) |
|---|---|---|---|
| $z \sim 1$ quasars | 70 | 85.0 ± 17.2 | 123.3 ± 27.6 |
| $-28.0 \leq M_i < -26.7$ | 17 | 186.0 ± 56.3 | 157.9 ± 40.9 |
| $-26.7 \leq M_i < -25.4$ | 18 | 88.6 ± 19.3 | 150.8 ± 29.4 |
| $-25.4 \leq M_i < -24.3$ | 17 | 85.6 ± 19.5 | 141.6 ± 20.9 |
| $-24.3 \leq M_i < -23.0$ | 18 | 22.9 ± 12.3 | 45.7 ± 10.6 |

Here we carry out a linear regression analysis between the measured radio-luminosity ($L_{1.5\,\text{GHz}}$) and the absolute *i*-band magnitude ($M_i$), for each of the 70 RQQs. The coefficient of determination, $R^2$, quantifies the goodness-of-fit on a scale from 0 to 1. Its value of 0.04 indicates that the one-to-one relation between the two intrinsic properties is poor. The translation of the best-fit line into log–linear space is presented in Fig. 5, and overplotted are the median values of the sample in four $M_i$ bins. Note that these median luminosities include the negative values derived for RQQs that had a negative flux-density extracted from the radio image. We test the correlation via the generalised Kendall rank statistic, $\tau$. This utilises the extra information provided by 2-$\sigma$ upper limits, for objects with non-detections. For Kendall correlation tests throughout this work we use the *bhkmethod* task (Feigelson & Nelson 1985; Isobe et al. 1986) from the STSDAS 'statistics' package, within the IRAF environment. The result of $\tau = -0.62$ and *p*-value $< 1.0 \times 10^{-4}$ (Table 2) indicates a strong anti-correlation between $L_{1.5\,\text{GHz}}$ and $M_i$. Due to optically brighter objects having a more negative $M_i$ value, this means that there is a significant correlation between the radio luminosity and the optical luminosity. In order to appreciate the influence of the upper limits on the results of the Kendall test, we repeat the test using only the RQQs that are detected above $2\sigma$ in the radio. This gives $\tau = -0.41$ and *p*-value $= 8.6 \times 10^{-2}$. Since we take *p*-value $\leq 1.0 \times 10^{-2}$ to indicate statistical significance, we cannot claim a significant anti-correlation for this subsample.

White et al. (2015) performed a Pearson correlation-test using their RQQ sample and found $r = -0.4$ and (median) *p*-value $= 10^{-4}$. Carrying out the same test for our sample gives $r = -0.1$ and *p*-value $= 0.3$. The reason for the weaker trend here may in part be due to observational effects present in the sample used by White et al. (2015), such as higher accretion rates observed at redshifts beyond $z \sim 1$ (Richards et al. 2006; Croom et al. 2009). To explore this further, deeper radio data on optically-selected quasars spanning a larger redshift range are needed, so that the correlation between radio and optical luminosity can be studied as a function of redshift.

## 6 ANALYSIS OF THE FIR EMISSION

Using the radio information described in Section 5, and FIR photometry already in hand, in this section we determine the relative contributions of black-hole accretion and star-formation processes to the total radio emission. However, rather than inferring the SFR from estimates of the stellar mass (as done by White et al. 2015), we fit a grey-body spectrum to the FIR emission to obtain a more-direct SFR measure, as detailed below.

### 6.1 Multi-band fitting in the far-infrared

Far-infrared emission is produced by the cool dust associated with star formation, and therefore acts as a good tracer of the SFR. This emission is typically approximated by a grey-body spectrum, which is the usual black-body function given by Planck's law,

$$B_\nu(\nu, T) = \frac{2h\nu^3}{c^2} \frac{1}{e^{\frac{h\nu}{k_B T}} - 1}, \tag{1}$$

multiplied by a frequency-dependent emissivity term, $Q_\nu = Q_0(\nu/\nu_0)^\beta$ (Hildebrand 1983). $B_\nu$ is the object's brightness at frequency $\nu$; $T$ is the dust temperature; $h$ is Planck's constant; $c$ is the speed of light; $k_B$ is Boltzmann's constant; and $\beta$ is the dust emissivity. The emissivity is usually in the range $1.0 < \beta < 2.0$, and encodes details about the dust grains, such as their shape, size, and chemical composition. As this leads to a degeneracy with temperature, we simplify the fitting by using a fixed value: $\beta = 1.8$. This is the weighted-mean $\beta$ used by Hardcastle et al. (2013) and Smith et al. (2013), whose samples consist of galaxies selected from *Herschel*-ATLAS (Astrophysical Terahertz Large Area Survey; Eales et al. 2010) at $z < 1$. Smith et al. (2013) also show that a simple, single-component, isothermal model performs equally well as a two-component model – the latter being used to simultaneously fit the FIR emission from a cold-dust population and a warm-dust population (Chini et al. 1986; de Jong & Brink 1987). Thus, with $\beta$ fixed and the (single) temperature allowed to vary between 10 and 60 K, the only other free parameter we use in the grey-body fitting code is the normalisation. Note that *K*-corrections are not an issue for this sample, as all of the quasars are at a similar redshift (Section 2).

We begin by using all of the available FIR photometry, as measured from PACS and SPIRE maps at 70, 160, 250, 350 and 500 μm (Section 3.2). However, if we are to determine the SFR from the FIR luminosity (through integration of the fitted spectrum, over 8–1000 μm; Sanders & Mirabel 1996), then we must be sure that this emission is not contaminated by dust-heating of AGN origin. Such a contribution is a concern for the 70 μm band at $z \sim 1$, as this corresponds to 35 μm in the rest frame. Here the AGN is heating its (putative) dusty torus, and this energy is re-radiated in the mid-infrared. As the AGN component of the mid- to far-infrared emission appears to become negligible beyond rest-frame 60 μm (e.g. Hatziminaoglou et al. 2010; Ivison et al. 2010; Feltre et al. 2014; Leipski et al. 2014), the longer-wavelength bands should be a reliable measure of the cooler dust, heated by star-formation processes.

As shown in Fig. 6, for a single-temperature model, the contribution of mid-infrared emission to the 70-μm flux biases the fit of the FIR emission to high temperatures. Without the constraint imposed by this band, the grey-body spectrum moves to longer wavelengths, where it is better able to describe the photometry (accompanied by a lower reduced-$\chi^2$ value). Therefore we proceed in fitting the FIR emission using only the 160, 250, 350 and 500 μm data, and note the improvement that this makes to the resulting

**Table 2.** Results of Kendall rank correlation tests, performed between different pairings of properties: total radio luminosity ($L_{1.5\,\mathrm{GHz}}$), absolute $i$-band magnitude ($M_i$), black-hole mass ($M_{\mathrm{BH}}$), SFR derived from the integrated-FIR luminosity [SFR($L_{\mathrm{FIR}}$)] using multi-band and single-band fitting (Fig. 9), star formation-related radio luminosity ($L_{1.5\,\mathrm{GHz,SF}}$), and accretion-related radio luminosity ($L_{1.5\,\mathrm{GHz,acc}}$). See Section 7.1 for the determination of $L_{1.5\,\mathrm{GHz,SF}}$ and $L_{1.5\,\mathrm{GHz,acc}}$. The Kendall statistic ($\tau$) quantifies the degree of correlation, with $\tau = 1$ indicating a strong correlation and $\tau = -1$ indicating a strong anti-correlation. The significance of the test, with respect to the null hypothesis that there is no correlation between the two properties concerned ($\tau = 0$), is given by the $p$-value (where we interpret $p$-value $\leq 1.0 \times 10^{-2}$ as being statistically significant). We exploit the ability of the Kendall test to accommodate upper or lower limits when considering the whole sample of RQQs, and not just those that are detected in the radio and/or the FIR (as appropriate). When considering the latter (i.e. a subset of the sample, as indicated under the column 'Description of objects used'), only measured values are used in the test. For other tests, the inability of the Kendall test to accommodate *both* upper and lower limits in a single variable means that we need to 'fix' a (minimum) number of limits, in order for the variable to consist of only detected values and *one* type of limit. (Where this is done is indicated under the column 'Description of objects used', with 'fixed limit' used to describe a limit value that is treated in the same way as a detected value.) Most objects have a value of $L_{1.5\,\mathrm{GHz,SF}}/L_{1.5\,\mathrm{GHz}}$ that is either fully constrained (in the case of radio-detected, FIR-detected RQQs) or partially constrained (in the case of RQQs detected in *only* the radio or the FIR). For the remaining objects, undetected in both the radio and the FIR, the fraction is unconstrained (Fig. 12), and so these objects are associated with both upper and lower limits in $L_{1.5\,\mathrm{GHz,SF}}/L_{1.5\,\mathrm{GHz}}$. Therefore, a Kendall test between $L_{1.5\,\mathrm{GHz,SF}}/L_{1.5\,\mathrm{GHz}}$ and $M_i$ is performed twice for the whole sample: once using the upper limits in $L_{1.5\,\mathrm{GHz,SF}}/L_{1.5\,\mathrm{GHz}}$, and once using the lower limits. (The fully- and partially-constrained values of $L_{1.5\,\mathrm{GHz,SF}}/L_{1.5\,\mathrm{GHz}}$ remain the same for these two Kendall tests.)

| Description of objects used | No. of objects | First property | Second property | Kendall statistic, $\tau$ | $p$-value |
|---|---|---|---|---|---|
| Whole sample | 70 | $L_{1.5\,\mathrm{GHz}}$ | $M_i$ | -0.62 | $< 1.0 \times 10^{-4}$ |
| Radio-detected RQQs | 35 | $L_{1.5\,\mathrm{GHz}}$ | $M_i$ | -0.41 | $8.6 \times 10^{-2}$ |
| Whole sample | 70 | SFR($L_{\mathrm{FIR}}$), multi-band | $M_{\mathrm{BH}}$ | 0.36 | $2.9 \times 10^{-2}$ |
| FIR-detected RQQs | 48 | SFR($L_{\mathrm{FIR}}$), multi-band | $M_{\mathrm{BH}}$ | 0.38 | $2.5 \times 10^{-2}$ |
| Whole sample | 70 | SFR($L_{\mathrm{FIR}}$), single-band | $M_{\mathrm{BH}}$ | 0.16 | $3.0 \times 10^{-1}$ |
| FIR-detected RQQs | 48 | SFR($L_{\mathrm{FIR}}$), single-band | $M_{\mathrm{BH}}$ | 0.05 | $8.0 \times 10^{-1}$ |
| Whole sample (using 9 fixed limits) | 70 | $L_{1.5\,\mathrm{GHz,acc}}$ | $L_{1.5\,\mathrm{GHz,SF}}$ | 0.37 | $6.1 \times 10^{-3}$ |
| Radio-detected, FIR-detected RQQs | 26 | $L_{1.5\,\mathrm{GHz,acc}}$ | $L_{1.5\,\mathrm{GHz,SF}}$ | 0.28 | $3.2 \times 10^{-1}$ |
| Whole sample | 70 | $L_{1.5\,\mathrm{GHz,SF}}$ | $M_i$ | -0.47 | $2.8 \times 10^{-3}$ |
| FIR-detected RQQs | 48 | $L_{1.5\,\mathrm{GHz,SF}}$ | $M_i$ | -0.33 | $9.5 \times 10^{-2}$ |
| Whole sample (using 9 fixed limits) | 70 | $L_{1.5\,\mathrm{GHz,acc}}$ | $M_i$ | -0.59 | $< 1.0 \times 10^{-4}$ |
| Radio-detected, FIR-detected RQQs | 26 | $L_{1.5\,\mathrm{GHz,acc}}$ | $M_i$ | -0.63 | $2.3 \times 10^{-2}$ |
| Whole sample (using 22 fixed limits) | 70 | $L_{1.5\,\mathrm{GHz,SF}}/L_{1.5\,\mathrm{GHz}}$, upper limits | $M_i$ | 0.28 | $1.8 \times 10^{-2}$ |
| Whole sample (using 9 fixed limits) | 70 | $L_{1.5\,\mathrm{GHz,SF}}/L_{1.5\,\mathrm{GHz}}$, lower limits | $M_i$ | 0.32 | $2.5 \times 10^{-3}$ |
| Radio-detected, FIR-detected RQQs | 26 | $L_{1.5\,\mathrm{GHz,SF}}/L_{1.5\,\mathrm{GHz}}$ | $M_i$ | 0.63 | $2.3 \times 10^{-2}$ |

reduced-$\chi^2$ values. The integrated-FIR luminosities calculated using these four bands are presented in Table 3.

### 6.2 Single-band fitting in the far-infrared

For comparison, we also carry out a simplified fit to the FIR emission, using a grey-body spectrum that is constrained by a single band: 250 μm. This band is chosen because, for this sample, it measures the emission close to the grey body's peak, which is expected to be at ~100 μm in the rest frame (Guiderdoni et al. 1998). Near the peak, any temperature variations lead to a minimal difference in the measured 250 μm flux, and so this can be a good indicator of the bolometric luminosity if the temperatures are similar. Technically the 160 μm band measures flux even closer to the peak, but this is on the shorter-wavelength side of the peak and so is more likely to suffer from AGN contamination. The 250 μm band therefore offers the best compromise.

For the single-band fitting, we use a fixed temperature and just fit for the normalisation ($\alpha$). To decide the most appropriate temperature to use, we inspect the current distribution for the sample (attained through multi-band fitting) and calculate the median temperature. This is 24.5 K. For comparison, Smith et al. (2013) used a fixed temperature of 23.5 K whilst Hardcastle et al. (2013) found that 20 K is the best-fitting global temperature for their sample. This suggests that the temperature of the dust present in this sample is similar to the general low-redshift galaxy population, and also to obscured radio galaxies. By using the median instead of the mean value, our results are less affected by outliers amongst the RQQs.

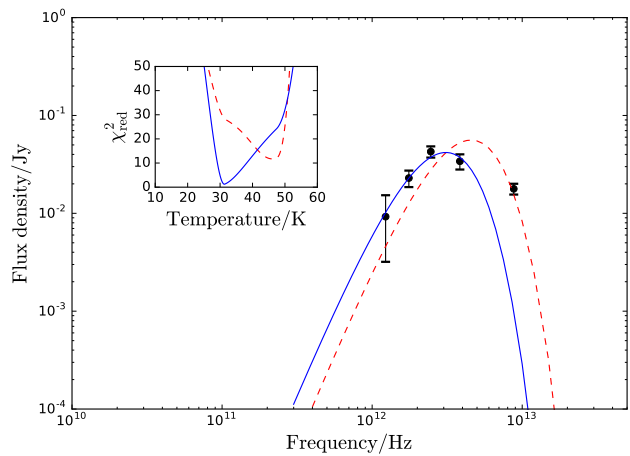

**Figure 6.** An example of fitting the far-infrared emission where all five PACS/SPIRE bands are used (red, dashed line). When the 70 μm flux is not included, there is an improvement in the fit, judged by the lower reduced-$\chi^2$ value (inset plot) and better overlap of the grey-body spectrum (blue, solid line) with the photometry (black datapoints).

**Table 3.** FIR measurements, and derived luminosities, for the 70 RQQs. The confusion limits for the 160, 250, 350 and 500μm bands are 0.68 (Magnelli et al. 2013), 5.8, 6.3 and 6.8 mJy beam$^{-1}$ (Nguyen et al. 2010), respectively. The resolution at 70μm is such that the observations are well above the predicted confusion level of 2.26 μJy beam$^{-1}$ (Lagache et al. 2003). The integrated-FIR luminosities presented, $L_{FIR}$, are determined using data at 160, 250, 350 and 500μm.

| R.A. (hms) | Dec. (dms) | $S_{70}$ (mJy) | $\sigma_{70}$ (mJy) | $S_{160}$ (mJy) | $\sigma_{160}$ (mJy) | $S_{250}$ (mJy) | $\sigma_{250}$ (mJy) | $S_{350}$ (mJy) | $\sigma_{350}$ (mJy) | $S_{500}$ (mJy) | $\sigma_{500}$ (mJy) | z | $L_{FIR}$ (W) | $L_{1.25\mu m}$ (W Hz$^{-1}$) |
|---|---|---|---|---|---|---|---|---|---|---|---|---|---|---|
| 00:31:46.07 | +13:46:30.0 | 2.20 | 2.63 | 7.85 | 4.09 | 8.45 | 5.95 | 3.82 | 4.98 | -0.53 | 7.38 | 1.01 | 6.85 × 10$^{37}$ | 2.24 × 10$^{25}$ |
| 02:35:40.90 | +00:10:39.2 | 4.81 | 2.34 | 18.20 | 4.93 | 6.97 | 7.01 | -5.50 | 5.70 | 0.31 | 9.06 | 0.95 | 2.85 × 10$^{38}$ | 1.64 × 10$^{25}$ |
| 07:38:02.35 | +38:31:16.5 | 35.02 | 2.46 | 64.91 | 4.67 | 104.45 | 6.32 | 70.82 | 5.15 | 32.79 | 7.39 | 1.02 | 6.93 × 10$^{38}$ | 2.85 × 10$^{26}$ |
| 07:47:29.16 | +43:46:07.8 | 3.46 | 2.56 | 8.21 | 4.76 | -1.70 | 6.36 | 10.25 | 5.11 | 6.73 | 6.26 | 1.09 | 3.94 × 10$^{37}$ | -5.20 × 10$^{24}$ |
| 07:50:58.20 | +42:16:16.9 | 2.84 | 2.28 | 5.62 | 4.98 | 3.24 | 6.45 | 16.42 | 4.90 | 15.45 | 6.46 | 0.94 | 5.67 × 10$^{37}$ | 7.44 × 10$^{24}$ |
| 07:52:22.89 | +27:38:23.0 | 20.85 | 2.65 | 14.18 | 5.32 | 8.92 | 6.17 | 1.09 | 4.97 | 3.79 | 7.17 | 1.06 | 1.80 × 10$^{38}$ | 2.60 × 10$^{25}$ |
| 07:53:39.81 | +25:01:37.8 | 6.87 | 2.34 | 11.16 | 5.18 | 6.36 | 8.84 | 15.31 | 7.30 | 3.41 | 8.58 | 0.94 | 7.51 × 10$^{37}$ | 1.48 × 10$^{25}$ |
| 08:22:29.76 | +44:27:05.2 | 8.99 | 2.39 | 4.73 | 5.81 | 0.93 | 7.02 | -7.16 | 8.04 | 4.60 | 8.30 | 1.06 | 3.55 × 10$^{37}$ | 2.71 × 10$^{24}$ |
| 08:31:15.86 | +42:33:16.5 | 16.87 | 2.54 | 48.53 | 5.45 | 58.69 | 5.52 | 39.14 | 5.66 | 10.96 | 6.96 | 0.93 | 3.65 × 10$^{38}$ | 1.33 × 10$^{26}$ |
| 08:47:23.64 | +01:10:10.3 | 3.75 | 2.30 | 13.86 | 5.74 | 17.24 | 7.69 | 18.07 | 5.11 | 5.80 | 7.17 | 1.08 | 1.52 × 10$^{38}$ | 5.25 × 10$^{25}$ |
| 09:01:53.42 | +06:59:15.3 | 7.61 | 2.49 | 15.07 | 4.77 | 25.98 | 6.50 | 31.90 | 5.55 | 25.57 | 7.58 | 1.08 | 2.22 × 10$^{38}$ | 7.92 × 10$^{25}$ |
| 09:12:16.87 | +42:03:14.3 | 2.82 | 2.65 | 14.20 | 4.67 | 10.48 | 6.95 | 16.83 | 5.32 | -0.70 | 7.25 | 1.08 | 1.31 × 10$^{38}$ | 3.17 × 10$^{25}$ |
| 09:22:57.86 | +44:46:52.1 | 12.29 | 2.74 | 29.01 | 5.75 | 32.54 | 5.75 | 33.80 | 5.36 | 12.81 | 7.30 | 1.08 | 3.02 × 10$^{38}$ | 9.83 × 10$^{25}$ |
| 09:27:53.52 | +05:36:36.8 | 8.83 | 2.31 | 21.11 | 5.67 | 12.90 | 7.69 | 9.53 | 6.29 | 16.33 | 9.04 | 1.06 | 1.96 × 10$^{38}$ | 3.79 × 10$^{25}$ |
| 09:28:29.78 | +50:48:36.3 | 2.42 | 2.73 | 8.72 | 5.31 | -5.27 | 6.54 | 5.71 | 5.22 | 0.72 | 7.48 | 1.03 | 7.95 × 10$^{37}$ | -1.40 × 10$^{25}$ |
| 09:30:23.28 | +40:31:11.1 | 6.05 | 2.49 | 19.05 | 5.20 | 21.71 | 6.58 | 11.73 | 5.27 | 9.06 | 6.59 | 1.10 | 2.07 × 10$^{38}$ | 6.80 × 10$^{25}$ |
| 09:33:03.48 | +46:04:39.8 | 6.26 | 2.33 | 15.88 | 4.86 | 22.10 | 6.16 | 19.78 | 5.43 | 14.35 | 7.75 | 1.09 | 1.92 × 10$^{38}$ | 6.83 × 10$^{25}$ |
| 09:37:59.35 | +54:24:27.2 | 6.48 | 2.51 | 19.99 | 5.29 | 62.29 | 6.18 | 80.68 | 5.31 | 70.17 | 8.41 | 1.07 | 4.79 × 10$^{38}$ | 1.85 × 10$^{26}$ |
| 09:48:11.85 | +55:17:26.4 | 3.62 | 3.14 | 7.10 | 5.26 | 14.23 | 6.01 | 14.44 | 5.09 | 1.56 | 6.79 | 1.03 | 9.35 × 10$^{37}$ | 3.97 × 10$^{25}$ |
| 10:07:30.48 | +05:09:42.0 | 13.32 | 2.19 | 56.86 | 5.37 | 68.67 | 6.36 | 40.09 | 5.40 | 8.79 | 5.13 | 0.92 | 4.17 × 10$^{38}$ | 1.52 × 10$^{26}$ |
| 10:08:35.83 | +51:39:27.9 | 5.60 | 2.50 | 12.09 | 5.13 | 11.92 | 5.95 | 7.95 | 4.88 | 8.23 | 7.01 | 1.08 | 1.23 × 10$^{38}$ | 3.65 × 10$^{25}$ |
| 10:09:06.33 | +02:35:55.4 | 2.72 | 2.71 | 4.24 | 2.65 | 21.46 | 6.24 | 9.63 | 4.04 | 5.42 | 7.89 | 1.10 | 1.10 × 10$^{38}$ | 6.76 × 10$^{25}$ |
| 10:20:05.97 | +03:33:08.4 | 8.33 | 2.70 | 13.11 | 5.09 | 19.08 | 6.16 | 19.84 | 4.98 | 16.93 | 6.59 | 0.94 | 1.16 × 10$^{38}$ | 4.37 × 10$^{25}$ |
| 10:21:11.56 | +61:14:15.2 | 2.94 | 2.42 | 5.48 | 5.05 | 10.44 | 6.02 | -1.69 | 5.75 | 3.01 | 6.70 | 0.93 | 4.71 × 10$^{37}$ | 2.37 × 10$^{25}$ |
| 10:23:49.39 | +52:21:51.2 | 9.41 | 5.63 | 21.14 | 5.63 | 42.85 | 6.52 | 36.92 | 5.74 | 13.39 | 6.61 | 0.96 | 2.30 × 10$^{38}$ | 1.02 × 10$^{26}$ |
| 10:33:47.30 | +09:40:39.0 | 13.46 | 2.71 | 19.63 | 5.01 | 31.29 | 5.98 | 25.15 | 5.21 | 19.31 | 7.58 | 1.03 | 2.21 × 10$^{38}$ | 8.63 × 10$^{25}$ |
| 10:35:25.05 | +58:03:35.6 | 2.78 | 2.28 | 11.10 | 5.12 | 22.24 | 5.89 | 15.64 | 5.22 | 6.54 | 7.17 | 0.96 | 1.19 × 10$^{38}$ | 5.41 × 10$^{25}$ |
| 10:38:29.73 | +58:52:04.1 | 6.99 | 2.85 | 17.26 | 5.83 | 32.41 | 6.17 | 22.88 | 5.21 | 16.10 | 6.88 | 0.93 | 1.67 × 10$^{38}$ | 7.41 × 10$^{25}$ |
| 10:38:55.32 | +57:58:14.7 | 4.77 | 2.49 | 11.32 | 5.06 | 14.15 | 5.04 | 12.82 | 4.88 | 9.87 | 7.01 | 0.96 | 9.43 × 10$^{37}$ | 3.38 × 10$^{25}$ |
| 10:41:14.18 | +59:02:19.5 | 4.17 | 2.72 | 5.99 | 5.44 | 10.06 | 5.95 | 4.77 | 5.75 | 2.85 | 8.24 | 1.09 | 7.54 × 10$^{37}$ | 3.13 × 10$^{25}$ |
| 10:42:39.64 | +58:32:31.0 | 3.46 | 2.51 | 8.95 | 5.63 | 14.02 | 6.16 | 10.87 | 5.40 | 10.09 | 7.48 | 1.00 | 9.29 × 10$^{37}$ | 3.65 × 10$^{25}$ |
| 10:43:55.46 | +59:30:54.0 | 4.22 | 2.38 | 19.64 | 5.42 | 15.55 | 6.24 | 13.50 | 4.18 | 8.73 | 7.58 | 0.91 | 1.22 × 10$^{38}$ | 3.36 × 10$^{25}$ |
| 10:45:37.68 | +48:49:14.6 | 2.20 | 2.47 | 18.13 | 5.46 | 19.36 | 5.93 | 29.38 | 4.95 | 12.71 | 5.95 | 0.94 | 1.40 × 10$^{38}$ | 4.50 × 10$^{25}$ |
| 10:46:59.37 | +57:30:55.8 | 3.22 | 2.46 | 5.43 | 5.60 | 5.75 | 6.84 | 9.47 | 5.74 | 11.47 | 6.46 | 1.03 | 5.16 × 10$^{37}$ | 1.58 × 10$^{25}$ |
| 10:48:59.66 | +56:56:48.6 | 6.24 | 2.99 | 17.40 | 5.49 | 13.01 | 6.27 | 25.35 | 5.09 | 14.18 | 6.26 | 1.01 | 1.38 × 10$^{38}$ | 3.49 × 10$^{25}$ |

**Table 3.** *Continued* – FIR measurements, and derived luminosities, for the 70 RQQs. The confusion limits for the 160, 250, 350 and 500 μm bands are 0.68 (Magnelli et al. 2013), 5.8, 6.3 and 6.8 mJy beam$^{-1}$ (Nguyen et al. 2010), respectively. The resolution at 70 μm is such that the observations are well above the predicted confusion level of 2.26 μJy beam$^{-1}$ (Lagache et al. 2003). The integrated-FIR luminosities presented, $L_{\rm FIR}$, are determined using data at 160, 250, 350 and 500 μm.

| R.A. (hms) | Dec. (dms) | $S_{70}$ (mJy) | $\sigma_{70}$ (mJy) | $S_{160}$ (mJy) | $\sigma_{160}$ (mJy) | $S_{250}$ (mJy) | $\sigma_{250}$ (mJy) | $S_{350}$ (mJy) | $\sigma_{350}$ (mJy) | $S_{500}$ (mJy) | $\sigma_{500}$ (mJy) | $z$ | $L_{\rm FIR}$ (W) | $L_{125\mu m}$ (W Hz$^{-1}$) |
|---|---|---|---|---|---|---|---|---|---|---|---|---|---|---|
| 10:49:30.45 | +59:20:32.7 | 8.92 | 3.43 | 12.72 | 4.65 | 10.05 | 6.21 | 6.16 | 5.09 | 4.69 | 7.38 | 1.01 | $1.07 \times 10^{38}$ | $2.68 \times 10^{25}$ |
| 10:49:35.76 | +55:49:50.5 | 6.06 | 2.56 | 9.72 | 5.22 | 14.35 | 6.24 | 19.98 | 5.43 | 25.09 | 6.88 | 1.06 | $1.24 \times 10^{38}$ | $4.17 \times 10^{25}$ |
| 10:54:08.88 | +04:26:50.4 | 9.20 | 2.14 | 35.22 | 5.62 | 38.59 | 5.44 | 34.75 | 4.78 | 17.26 | 6.21 | 1.08 | $3.68 \times 10^{38}$ | $1.18 \times 10^{26}$ |
| 11:23:17.49 | +05:18:03.9 | 20.88 | 2.60 | 16.75 | 5.01 | 18.04 | 6.32 | 4.87 | 5.91 | 2.25 | 8.24 | 1.00 | $1.46 \times 10^{38}$ | $4.72 \times 10^{25}$ |
| 11:50:27.21 | +66:58:48.1 | 11.34 | 2.37 | 9.22 | 4.76 | 15.84 | 5.04 | 13.73 | 4.18 | 4.91 | 5.13 | 1.04 | $1.08 \times 10^{38}$ | $4.43 \times 10^{25}$ |
| 12:28:32.92 | +60:37:35.1 | 2.48 | 2.51 | 6.13 | 5.84 | 6.62 | 5.34 | 9.46 | 4.87 | 5.73 | 6.88 | 1.04 | $5.81 \times 10^{37}$ | $1.87 \times 10^{25}$ |
| 12:30:59.71 | +10:16:24.5 | 1.72 | 2.31 | 3.94 | 4.76 | 18.87 | 6.32 | 7.82 | 6.31 | 6.15 | 5.13 | 1.06 | $9.40 \times 10^{37}$ | $5.49 \times 10^{25}$ |
| 12:56:59.90 | +04:27:34.6 | 53.24 | 2.51 | 57.65 | 5.86 | 65.53 | 6.46 | 42.30 | 5.40 | 21.01 | 7.01 | 1.03 | $5.33 \times 10^{38}$ | $1.80 \times 10^{26}$ |
| 13:29:57.14 | +54:05:06.0 | 27.86 | 2.78 | 19.38 | 5.11 | 21.14 | 6.27 | 1.71 | 5.41 | 3.26 | 7.00 | 0.95 | $1.58 \times 10^{38}$ | $4.98 \times 10^{25}$ |
| 13:37:13.00 | +61:07:49.0 | 3.07 | 2.33 | 6.44 | 6.01 | 7.20 | 6.54 | 7.40 | 5.14 | 6.86 | 8.04 | 0.93 | $4.78 \times 10^{37}$ | $1.62 \times 10^{25}$ |
| 13:37:33.28 | +59:06:22.8 | 16.15 | 2.23 | 14.01 | 6.31 | 25.71 | 5.81 | 22.39 | 5.00 | 18.97 | 6.72 | 1.09 | $1.98 \times 10^{38}$ | $7.91 \times 10^{25}$ |
| 13:58:23.97 | +02:13:44.0 | 8.16 | 2.49 | 8.46 | 5.02 | 6.67 | 6.02 | −0.39 | 5.21 | −1.09 | 6.63 | 0.96 | $8.99 \times 10^{37}$ | $1.60 \times 10^{25}$ |
| 14:21:24.67 | +42:30:03.1 | 9.86 | 2.43 | 21.44 | 5.99 | 31.63 | 6.17 | 38.37 | 5.02 | 31.02 | 7.24 | 1.00 | $2.26 \times 10^{38}$ | $8.27 \times 10^{25}$ |
| 14:28:17.30 | +50:27:12.7 | 11.72 | 2.43 | 10.58 | 5.46 | 7.86 | 6.24 | 12.87 | 4.99 | 3.70 | 7.71 | 1.01 | $8.26 \times 10^{37}$ | $2.11 \times 10^{25}$ |
| 14:55:03.45 | +01:42:09.2 | 4.36 | 2.42 | 21.94 | 5.40 | 40.95 | 6.08 | 56.54 | 5.61 | 21.51 | 7.84 | 1.05 | $3.17 \times 10^{38}$ | $1.18 \times 10^{26}$ |
| 14:55:06.09 | +56:29:35.6 | 9.66 | 2.76 | 25.06 | 5.97 | 19.20 | 5.89 | 24.28 | 4.99 | 7.83 | 5.55 | 1.04 | $2.09 \times 10^{38}$ | $5.40 \times 10^{25}$ |
| 15:15:20.54 | +00:47:39.4 | 14.47 | 2.39 | 36.68 | 5.46 | 52.51 | 6.84 | 39.97 | 6.24 | 25.63 | 8.90 | 0.95 | $3.17 \times 10^{38}$ | $1.24 \times 10^{26}$ |
| 15:19:21.84 | +53:58:42.2 | 2.91 | 2.63 | 4.91 | 4.44 | 2.98 | 6.50 | 0.86 | 5.88 | −2.00 | 6.36 | 1.03 | $4.74 \times 10^{37}$ | $8.20 \times 10^{24}$ |
| 15:54:36.24 | +32:04:08.5 | 5.58 | 2.37 | 11.47 | 5.87 | 23.63 | 6.06 | 24.20 | 5.04 | 13.01 | 7.59 | 1.06 | $1.67 \times 10^{38}$ | $6.89 \times 10^{25}$ |
| 15:56:50.40 | +39:45:42.8 | 16.32 | 2.63 | 70.04 | 5.07 | 70.57 | 6.21 | 64.89 | 5.62 | 23.46 | 7.05 | 0.94 | $5.14 \times 10^{38}$ | $1.64 \times 10^{26}$ |
| 16:32:25.56 | +41:18:52.4 | 6.04 | 2.63 | 12.29 | 5.58 | 12.62 | 6.54 | 16.48 | 4.88 | 8.54 | 7.53 | 0.91 | $8.35 \times 10^{37}$ | $2.73 \times 10^{25}$ |
| 16:33:06.12 | +40:17:47.5 | 5.10 | 2.49 | 9.30 | 5.69 | 19.34 | 5.87 | 28.82 | 6.24 | 15.34 | 6.21 | 0.97 | $1.28 \times 10^{38}$ | $4.80 \times 10^{25}$ |
| 16:34:08.64 | +33:12:42.1 | 4.19 | 2.59 | 5.39 | 5.18 | 11.83 | 6.36 | 11.02 | 5.52 | −3.61 | 6.90 | 1.01 | $6.71 \times 10^{37}$ | $3.13 \times 10^{25}$ |
| 16:39:30.81 | +41:00:13.6 | 2.67 | 2.44 | 8.79 | 5.71 | 13.36 | 6.52 | 10.93 | 6.25 | 3.52 | 5.95 | 1.05 | $9.95 \times 10^{37}$ | $3.85 \times 10^{25}$ |
| 16:46:17.16 | +36:45:09.6 | 3.54 | 2.29 | 9.45 | 5.31 | 19.12 | 6.23 | 29.14 | 4.94 | 14.10 | 6.96 | 0.96 | $1.22 \times 10^{38}$ | $4.59 \times 10^{25}$ |
| 16:52:31.29 | +35:36:15.9 | 4.46 | 2.46 | 22.62 | 5.74 | 32.60 | 6.32 | 26.42 | 5.74 | 6.55 | 7.01 | 0.93 | $1.84 \times 10^{38}$ | $7.35 \times 10^{25}$ |
| 17:10:05.52 | +64:48:42.9 | 21.71 | 2.52 | 65.92 | 5.41 | 94.70 | 6.11 | 76.58 | 4.97 | 46.75 | 7.21 | 1.01 | $6.56 \times 10^{38}$ | $2.51 \times 10^{26}$ |
| 17:11:45.52 | +60:13:18.6 | 3.72 | 2.44 | 5.45 | 5.59 | 8.07 | 6.32 | 12.25 | 6.31 | 15.06 | 7.58 | 0.98 | $6.13 \times 10^{37}$ | $2.03 \times 10^{25}$ |
| 17:13:30.24 | +64:42:53.0 | 17.80 | 2.25 | 34.04 | 6.02 | 42.70 | 5.60 | 22.93 | 4.39 | 9.25 | 6.06 | 1.05 | $3.48 \times 10^{38}$ | $1.23 \times 10^{26}$ |
| 17:17:04.68 | +28:14:00.7 | 36.59 | 2.57 | 44.26 | 4.02 | 42.86 | 7.38 | 35.24 | 6.31 | 25.47 | 7.24 | 1.08 | $4.45 \times 10^{38}$ | $1.30 \times 10^{26}$ |
| 17:17:32.92 | +59:47:47.5 | 7.39 | 2.55 | 17.06 | 5.12 | 27.74 | 6.24 | 25.23 | 5.22 | 9.22 | 6.61 | 1.06 | $2.08 \times 10^{38}$ | $8.12 \times 10^{25}$ |
| 17:21:30.96 | +58:44:04.7 | 2.58 | 2.57 | 22.47 | 5.13 | 15.31 | 5.44 | 15.24 | 5.02 | 15.40 | 6.70 | 1.00 | $1.74 \times 10^{38}$ | $4.00 \times 10^{25}$ |
| 17:23:10.34 | +59:51:05.7 | 7.41 | 2.64 | 22.65 | 5.29 | 30.38 | 6.17 | 20.15 | 5.22 | 12.11 | 7.01 | 0.99 | $2.06 \times 10^{38}$ | $7.78 \times 10^{25}$ |
| 21:55:41.73 | +12:28:18.9 | 8.49 | 2.47 | 16.42 | 5.54 | 15.40 | 7.34 | 3.75 | 6.25 | −1.47 | 7.63 | 1.06 | $1.73 \times 10^{38}$ | $4.55 \times 10^{25}$ |
| 22:41:59.42 | +14:20:55.0 | 5.88 | 2.35 | 9.44 | 4.95 | 15.76 | 5.87 | 14.30 | 5.22 | 8.04 | 5.80 | 0.95 | $9.17 \times 10^{37}$ | $3.75 \times 10^{25}$ |

## 6.3 Determining the far-infrared luminosity

For both the multi-band and single-fitting cases, the FIR luminosity for each RQQ is calculated using the following integral:

$$L_{\rm FIR} = \frac{4\pi D_L^2}{1+z} \int \alpha Q_\nu B_\nu d\nu, \quad (2)$$

where $D_L$ is the luminosity distance, and the integration is performed over rest-frame wavelengths $8 < \lambda/\mu{\rm m} < 1000$ (Sanders & Mirabel 1996; Bell 2003).

## 7 INVESTIGATING THE LEVEL OF STAR FORMATION

Previously, Rosario et al. (2013) showed that most quasar hosts exhibit a level of star formation that is consistent with normal, star-forming galaxies. To investigate whether star-formation activity is sufficient to explain the total radio emission for this sample, we make use of the well-established far-infrared to radio correlation (FIRC, Section 1.1). This is followed by a comparison of two estimates of the SFR, using a method similar to that of White et al. (2015).

### 7.1 The far-infrared–radio correlation

The level of star formation in quasars can be traced via their FIR luminosity. Combining this with the FIRC allows us to determine how much radio emission is expected to be associated with this process, and therefore what fraction of the total radio emission must be due to accretion. A temperature-dependence in the FIRC has been identified by Smith et al. (2014), and so we use a measurement of the FIR emission that minimises any bias with respect to temperature. As already explained (Section 6.2), for this sample, the 250 μm flux-density is a reliable tracer that meets this criterion. We therefore use this measurement to construct a FIRC that is appropriate for these RQQs.

Following Smith et al. (2014), we use a dimensionless parameter, $q_\lambda$, to describe the FIRC:

$$q_\lambda = \log_{10}\left[\frac{L_\lambda}{L_{1.5\,{\rm GHz}}}\right], \quad (3)$$

where $\lambda$ is the rest-frame wavelength. Since we measure the FIR emission using the 250 μm band, and so directly probe the intrinsic 125-μm emission for $z \sim 1$ objects, we calculate the monochromatic luminosity at rest-frame 125 μm and use this for our FIRC parameter, as defined in Equation 4. The work of Smith et al. (2014) included investigation of the temperature dependence of $q_{100}$ and $q_{160}$ (which use the $K$-corrected flux-densities at 100 μm and 160 μm, respectively, for Equation 3). They present stacked values for these $q$ parameters, and we use the extremes of these values ($q_{160} = 2.4$ and $q_{100} = 2.9$) as a guide for what we may expect our $q_{125}$ values to be. (As the 125 μm rest-frame emission lies between emission at 100 μm and emission at 160 μm, it is expected that the behaviour of $q_{125}$ should be bounded by the behaviour of $q_{100}$ and $q_{160}$.) Note that the values 2.4 and 2.9 are taken from Figure 9 of Smith et al. (2014), and we simply use the midpoint value for the definition below:

$$q_{125} = \log_{10}\left[\frac{L_{125\,\mu{\rm m}}/({\rm W\,Hz^{-1}})}{L_{1.5\,{\rm GHz}}/({\rm W\,Hz^{-1}})}\right] = 2.65. \quad (4)$$

The positions of the RQQs with respect to the FIRC can be seen in Fig. 7. Purely star-forming galaxies are expected to lie between the dashed lines, but the majority of the sample are towards the right-hand side of the FIRC. This means that the total radio emission exceeds that expected from star formation alone. Hence, we have good evidence that the accretion process makes a significant contribution to the radio emission from RQQs. The current work is complementary to that of White et al. (2015), having different selection effects and biases. It also benefits from using individual measurements of $L_{125\,\mu{\rm m}}$ for each quasar, rather than relying on probability contours derived from assumed black-hole masses and empirical scaling relations.

To determine the level of the contribution by star formation and the AGN to the total radio luminosity, we use the FIRC ($q_{125} = 2.65$) to calculate the radio luminosity based on the value of $L_{125\,\mu{\rm m}}$ for each object. This gives the radio luminosity expected due to star-formation processes, $L_{1.5\,{\rm GHz, SF}}$. Subtracting this from the total radio luminosity then provides an estimate of the radio luminosity that is connected to accretion:

$$L_{1.5\,{\rm GHz, acc}} = L_{1.5\,{\rm GHz}} - L_{1.5\,{\rm GHz, SF}}. \quad (5)$$

This is carried out for each of the RQQs, including the one lying to the left-hand side of the solid line in Fig. 7. This object is likely to have a negligible contribution from accretion to the radio emission, and (by the above definition) have an unphysical value of $L_{1.5\,{\rm GHz, acc}}$. For cases where the RQQ is detected in the radio but undetected in the FIR, the upper limit in $L_{1.5\,{\rm GHz, SF}}$ leads to a lower limit in $L_{1.5\,{\rm GHz, acc}}$.

We then use the derived $L_{1.5\,{\rm GHz, acc}}$ values for the full sample to determine what fraction of the sample has AGN-dominated radio emission (defined as $L_{1.5\,{\rm GHz, acc}}/L_{1.5\,{\rm GHz}} > 0.5$), and the total accretion-related radio luminosity for all of the objects. This is repeated using the derived values of $L_{1.5\,{\rm GHz, acc}}$ for the RQQs detected in both the radio and the FIR, in combination with 2-$\sigma$ upper limits in $L_{1.5\,{\rm GHz, acc}}$ for the remainder of the sample. We then repeat this a second time, again using the derived values of $L_{1.5\,{\rm GHz, acc}}$ for the radio-detected, FIR-detected RQQs, but in combination with 2-$\sigma$ *lower* limits in $L_{1.5\,{\rm GHz, acc}}$ for the remaining RQQs. Doing so allows us to appreciate the influence of the non-detections (in the radio and/or the FIR) on the final values that we use to quantify accretion-related radio emission. For the whole sample, we find that the fraction of AGN-dominated RQQs is in the range 47–89 per cent, with the total accretion-related radio luminosity accounting for 60–83 per cent of the summed radio-luminosity (Table 4). Note that wherever derived values of $L_{1.5\,{\rm GHz, acc}}$ are used (as indicated under the 'Description of values used' column in Table 4), the summed radio-luminosity is also a derived value. Otherwise, for samples that include RQQs undetected in the radio, the summed radio-luminosity is an upper limit.

We then also consider, in turn, objects belonging to each of the four 'categories' of detection (defined by the level of their detection in the radio and the FIR, either above or below a 2-$\sigma$ threshold). The most robust of these are the radio-detected, FIR-detected RQQs, which make up over one third of the sample. 92 per cent of these objects have radio emission that is dominated by the AGN, with the accretion process accounting for 80 per cent of the total radio luminosity, summed across the 26 objects. This is comparable to the findings of Herrera Ruiz et al. (2016), who showed that AGN activity accounts for 50–75 per cent of the radio emission in three RQQs, detected using the Very Long Baseline Array (VLBA). In the case of the radio-detected, FIR-undetected RQQs, all nine of these objects are AGN-dominated. The fact that lower limits in $L_{1.5\,{\rm GHz, acc}}$ have been used means that the fraction of the summed radio-luminosity, that is related to accretion, is *at least* 90 per cent. (Note that for samples consisting only radio-detected RQQs

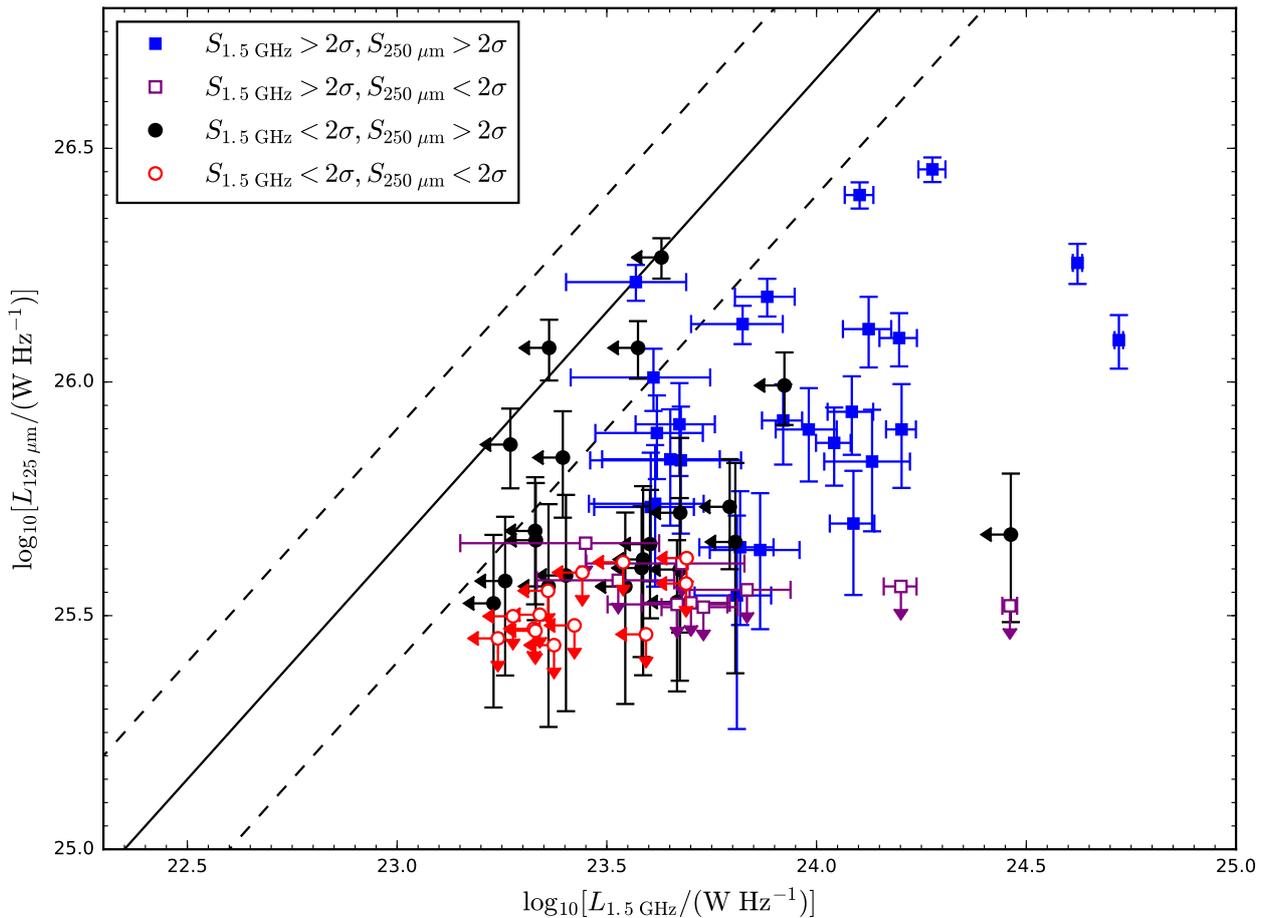

**Figure 7.** The monochromatic luminosity at rest-frame 125 μm, $L_{125\,\mu m}$, versus the radio luminosity, $L_{1.5\,GHz}$. Squares correspond to objects detected in the radio above $2\sigma$, and circles are those below this detection threshold. Note that objects with $\log_{10}[L_{125\,\mu m}] \lesssim 25.6$ are below $2\sigma$ at 250 μm (unfilled symbols). Arrows represent 2-$\sigma$ upper limits in $L_{1.5\,GHz}$ or $L_{125\,\mu m}$, for quasars undetected at the 2-$\sigma$ level in the JVLA images (horizontal arrows) or the 250 μm photometry (vertical arrows), respectively. The dashed lines are the lower and upper bounds on $q_{125}$, 2.4 and 2.9, respectively. The solid line corresponds to the midpoint value, $q_{125} = 2.65$ (Equation 4).

– whether or not they are also detected in the FIR – the summed radio-luminosity is a derived value.) For the 22 radio-undetected, FIR-detected RQQs, we use upper limits in $L_{1.5\,GHz,\,acc}$. This indicates that a maximum of 73 per cent of the objects in question have AGN-dominated radio emission, and that accretion accounts for $\leq 72$ per cent of the total, summed radio-luminosity. Lastly, we consider the remaining 13 RQQs. These are undetected in both the radio and the FIR and, as such, have a value of $L_{1.5\,GHz,\,acc}$ that is unconstrained. This means that it could lie anywhere between zero (the lower limit) and a value equal to the total radio luminosity, $L_{1.5\,GHz}$ (the upper limit). Therefore, depending on whether the lower or upper limits are used, the accretion-related fraction is the minimum or maximum possible value (0.0 or 1.0, respectively). We still present this subsample in Table 4 for completeness.

Taking our analysis a step further, we separately study the RQQs for which the total radio luminosity is in excess of the radio luminosity corresponding to $q_{125} = 2.15$ (given the measured 125 μm luminosity). This equates to the object lying over $2\sigma$ away from the FIRC, in the direction of more radio-luminous sources. Although all of these objects are clearly AGN-dominated, we are still interested in quantifying the accretion component of their radio emission. For selecting this subset of RQQs, we have approximated the error in the FIRC via the relative positions of the lower and upper bounds ($\Delta q_{125} = 0.25$, see Fig. 7). In addition, we calculate the error in the measured $q_{125}$ value ($\sigma_q$) for each RQQ, to ensure that we only consider the objects that are significantly offset from the FIRC. Depending on whether lower or upper limits in $L_{1.5\,GHz,\,acc}$ are used, this corresponds to 29–54 per cent of the full sample, and the summed accretion-related contribution to the total radio emission is at a level of 89–92 per cent (Table 4).

We note that, by performing a stacking analysis on their sample of star-forming galaxies, Smith et al. (2014) find no evidence for redshift evolution in the monochromatic FIRC out to $z \sim 0.4$. We expect there to be little evolution between their findings and our sample at $z \sim 1$, as this is still before the critical redshift at which inverse-Compton losses (off the cosmic microwave background) leads to suppression of the radio emission (Lacki & Thompson 2010). Although Magnelli et al. (2015) do find evidence for (moderate) evolution in the FIRC, the error that may be introduced by extrapolating from $z \sim 0.4$ to $z \sim 1$ is smaller than $\Delta q_{125}$. Therefore, our monochromatic FIRC ($q_{125} = 2.65$) is expected to still hold at $z \sim 1$, within the errors.

**Table 4.** The accretion-related contribution to the radio luminosity, across the sample of RQQs. An object with $L_{1.5\,\mathrm{GHz,acc}}/L_{1.5\,\mathrm{GHz}} > 0.5$ is described as being 'AGN-dominated'. 'Summed radio luminosity' refers to the summation of the total radio luminosity for each object ($\Sigma L_{1.5\,\mathrm{GHz}}$), and the 'fraction that is accretion-related' is given by $\Sigma L_{1.5\,\mathrm{GHz,acc}}/\Sigma L_{1.5\,\mathrm{GHz}}$. 'Upper' and 'lower' limits refer to the fraction of the radio emission that is related to accretion, taking into account cases where the object is undetected (i.e. $< 2\sigma$) in the radio and/or the FIR. Where the value of $q$ (Equation 4) is used to define a subset of the RQQ sample, the error in $q$ ($\sigma_q$) has been considered (i.e. we retain objects where $q + \sigma_q < 2.15$).

| Description of objects used | No. of objects | Fraction that are AGN-dominated | Summed radio luminosity (W Hz$^{-1}$) | Fraction of summed luminosity that is accretion-related |
|---|---|---|---|---|
| Whole sample (derived values of $L_{1.5\,\mathrm{GHz,acc}}$) | 70 | 0.80 | $3.82 \times 10^{25}$ | 0.74 |
| Whole sample (lower limits in $L_{1.5\,\mathrm{GHz,acc}}$) | 70 | $\geq 0.47$ | $\leq 5.28 \times 10^{25}$ | $\geq 0.60$ |
| Whole sample (upper limits in $L_{1.5\,\mathrm{GHz,acc}}$) | 70 | $\leq 0.89$ | $\leq 5.28 \times 10^{25}$ | $\leq 0.83$ |
| Radio-detected, FIR-detected objects (derived values) | 26 | 0.92 | $3.07 \times 10^{25}$ | 0.80 |
| Radio-detected, FIR-undetected objects (lower limits) | 9 | 1.00 | $7.76 \times 10^{24}$ | $\geq 0.90$ |
| Radio-undetected, FIR-detected objects (upper limits) | 22 | $\leq 0.73$ | $\leq 1.07 \times 10^{25}$ | $\leq 0.72$ |
| Radio-undetected, FIR-undetected objects (lower limits) | 13 | $\geq 0.00$ | $\leq 3.73 \times 10^{24}$ | $\geq 0$ |
| Radio-undetected, FIR-undetected objects (upper limits) | 13 | $\leq 1.00$ | $\leq 3.73 \times 10^{24}$ | $\leq 1.00$ |
| Objects right-wards of $q = 2.15$ (derived values) | 19 | 1.00 | $2.77 \times 10^{25}$ | 0.89 |
| Objects right-wards of $q = 2.15$ (lower limits) | 20 | 1.00 | $\leq 2.93 \times 10^{25}$ | $\geq 0.89$ |
| Objects right-wards of $q = 2.15$ (upper limits) | 38 | $\leq 1.00$ | $\leq 3.75 \times 10^{25}$ | $\leq 0.92$ |

### 7.2 Comparison of star-formation rates

Another method for investigating the origin of the radio emission is to compare two independent estimates of the SFR. Our first estimate is calculated directly from the radio luminosity, following the relation of Yun et al. (2001):

$$\mathrm{SFR}/(\mathrm{M}_\odot\,\mathrm{yr}^{-1}) = 5.9 \times 10^{-22} L_{1.4\,\mathrm{GHz}}/(\mathrm{W\,Hz}^{-1}). \quad (6)$$

This assumes that the total flux in the radio is due to star formation alone. The second SFR estimate is derived from the FIR luminosity, $L_{\mathrm{FIR}}$, as determined in the previous section. For this we use Equation 7, taken from Kennicutt (1998), which applies to starbursts:

$$\mathrm{SFR}/(\mathrm{M}_\odot\,\mathrm{yr}^{-1}) = 4.5 \times 10^{-44} L_{\mathrm{FIR}}/(\mathrm{erg\,s}^{-1}). \quad (7)$$

This is most relevant for our purposes, as the SFR in such objects is at a maximum, limited only by the amount of dense gas that is available. Alternatively, driven by galaxies with low SFRs, Bell (2003) suggest a two-power law relation between far-infrared and radio emission. However, this second power-law is only warranted at very low SFRs, which are well below the SFR that could explain any of the radio emission in this RQQ sample. In addition, although not all of the light from star formation may be reprocessed by the dust (leading to the SFR being underestimated when inferred from $L_{\mathrm{FIR}}$), Hayward et al. (2014) show that this effect is only significant for isolated, low-mass systems. For isolated, high-mass systems, $L_{\mathrm{FIR}}$ is a good tracer of the true SFR, whilst for merging systems (i.e. starburst and post-starburst phases), they find that Equation 7 leads to an *over*estimation of the SFR. This is due to the dust being heated by a large number of young stars (hundreds of Myr old), whilst the SFR relation is technically only applicable for starbursts with ages less than 100 Myr. Therefore, the SFR($L_{\mathrm{FIR}}$) values derived for our sample of RQQs are themselves upper limits.

In Fig. 8 we show the two SFR estimates against one another, for both the multi-band and single-band fitting cases. For quasars that are undetected in the 250 μm map (i.e. are below $2\sigma$), the 2-$\sigma$ upper limit for this measurement is used instead when determining $L_{\mathrm{FIR}}$. As such, the SFR derived from this is also a 2-$\sigma$ upper limit (vertical arrows in Fig. 8). Note that the results are very similar for the two types of fitting. This demonstrates that sampling close to the peak emission provides a good indication of the integrated far-infrared luminosity.

If the total radio emission from the sample could be explained entirely by star formation, then the datapoints should lie along a one-to-one relation (dashed lines in Fig. 8). We find that the majority of the RQQs have values of SFR($L_{1.5\,\mathrm{GHz}}$) that exceed this relation, suggesting that the radio emission is above that expected from star formation alone. Fig. 8 is effectively the same as Fig. 7 but with the conversion from observable quantities ($L_{1.5\,\mathrm{GHz}}$ and $L_{\mathrm{FIR}}$ – the latter closely-correlated with $L_{125\mu\mathrm{m}}$) to SFRs. As before, the accretion-connected radio emission in RQQs is clearly significant.

Next we investigate whether any correlation exists between SFR($L_{\mathrm{FIR}}$) and black-hole mass, $M_{\mathrm{BH}}$ (Section 3.1), for each of the 70 RQQs (Fig. 9). It may be expected that, because of the larger gas reservoirs associated with more-massive galaxies (and, implicitly, more-massive black holes), the amount of star formation increases with black-hole mass. However, only a very weak trend can be seen by eye between SFR($L_{\mathrm{FIR}}$) and $M_{\mathrm{BH}}$ (Fig. 9), whilst the multi-band fitting and single-band fitting cases lead to Kendall statistics of $\tau = 0.36$ ($p$-value $= 2.9 \times 10^{-2}$) and $\tau = 0.16$ ($p$-value $= 3.0 \times 10^{-1}$), respectively (Table 2, where upper limits have been used for all FIR non-detections). Note that neither of these results is statistically significant. For comparison, we also consider only the objects that are detected in the 250 μm map, so that no upper limits are used in the Kendall tests. These tests show that the trend is again non-existent for both the multi-band case ($\tau = 0.38$, $p$-value $= 2.5 \times 10^{-2}$) and for the SFR($L_{\mathrm{FIR}}$) values derived from single-band fitting ($\tau = 0.05$, $p$-value $= 8.0 \times 10^{-1}$). What is apparent, however, is the higher radio-detection ($>2\sigma$) rate for larger black-hole masses. This ranges from 33 per cent in the lowest-$M_{\mathrm{BH}}$ bin to 72 per cent in the highest-$M_{\mathrm{BH}}$ bin, and is consistent with the results of McLure & Jarvis (2004).

Fig. 9 also illustrates the difficulty of using sample properties to determine how various galaxy-evolution processes interact. One explanation for the lack of strong correlation between $M_{\mathrm{BH}}$ and SFR is that the typical gas mass is more dependent on halo mass and the age of the system, rather than how massive the galaxy is. The scatter in the $M_{\mathrm{BH}}$–stellar-mass relation is then compounded by the scatter in the stellar-mass–SFR relation. With larger black-hole mass not necessarily being connected to higher SFR, this may seem to contradict the star-formation mass sequence studied by numerous authors (e.g. Noeske et al. 2007; Whitaker et al. 2012; Johnston et al. 2015). However, note that this relation was derived from star-forming galaxies, for which the stellar mass rather than black-hole mass is used. We also assume that the host galaxies of RQQs behave in the same way as normal star-forming galaxies, as suggested by Rosario et al. (2013). Another consideration is that a larger black-hole accretion rate may restrict how much gas is available for star formation, but similarly, star-formation activity may place a limit on the volume of gas that can be accreted (cf. Bouché et al. 2010). In the absence of any CO observations to directly determine the abundance of gas, we can only study the relative contributions of star formation and accretion to the total radio emission.

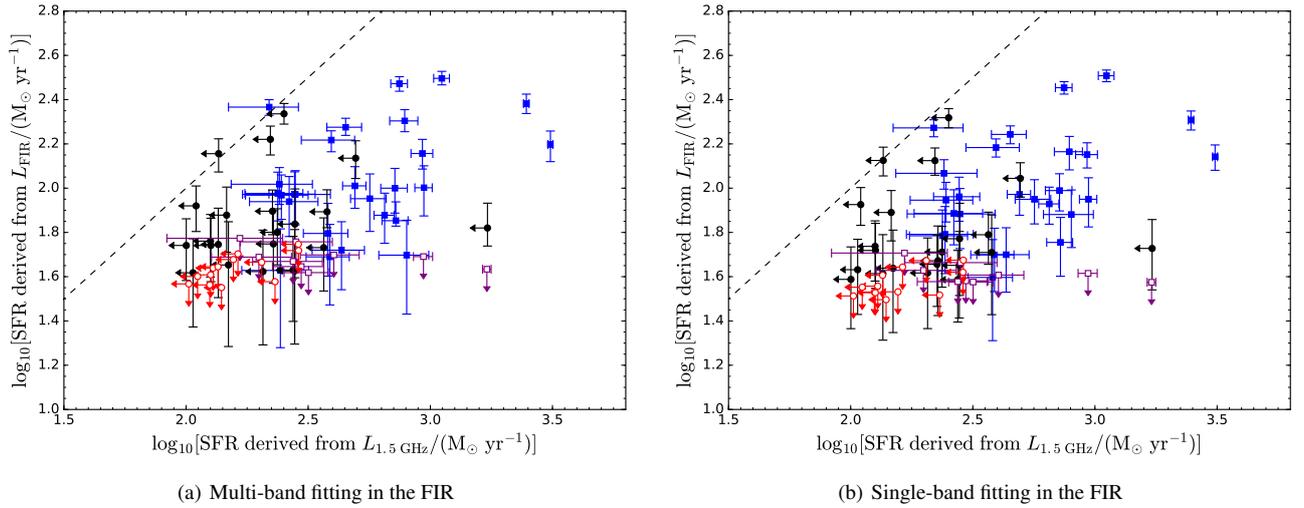

**Figure 8.** The SFR calculated from the integrated-FIR luminosity ($L_{\rm FIR}$, Section 6) against the SFR calculated from the extracted radio flux-density at 1.5 GHz. $L_{\rm FIR}$ is determined by using either (a) 4 bands to fit the grey-body spectrum, or (b) the 250 μm measurement only. Squares (blue and purple) correspond to objects detected in the radio above $2\sigma$, and circles (red and black) are those below this detection threshold. Unfilled symbols correspond to the FIR data being below $2\sigma$. Arrows represent 2-$\sigma$ upper limits in the SFR for quasars undetected at the 2-$\sigma$ level in the JVLA images (horizontal arrows) or the 250 μm photometry (vertical arrows). The dashed line indicates the one-to-one relation, if star formation accounts for all of the radio emission.

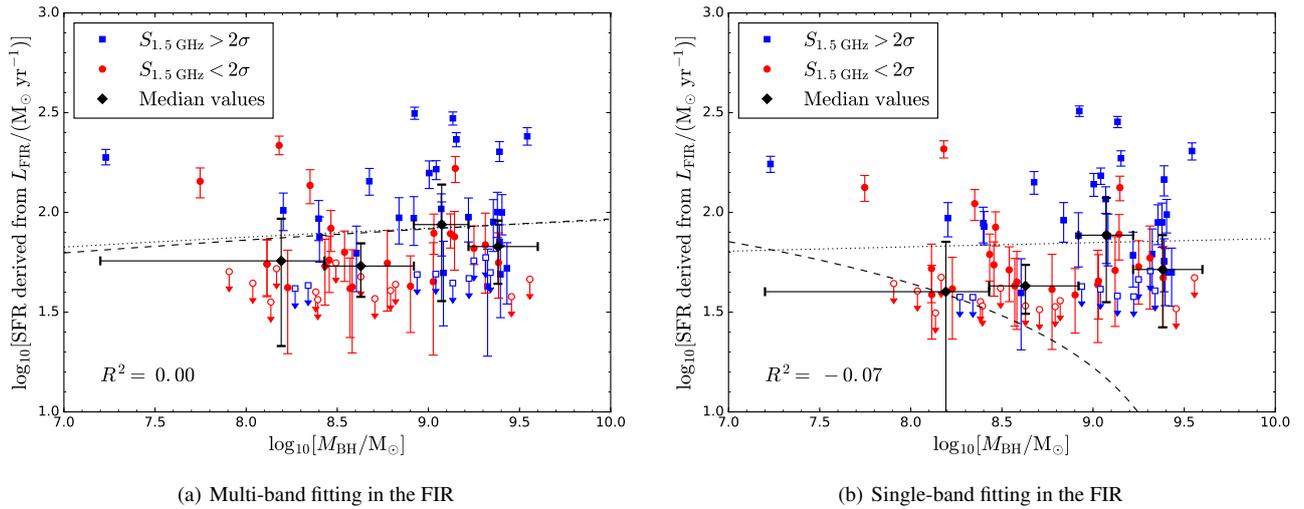

**Figure 9.** The SFR calculated from the integrated-FIR luminosity [SFR($L_{\rm FIR}$), Section 6] against the black-hole mass, $M_{\rm BH}$. $L_{\rm FIR}$ is determined by using either (a) 4 bands to fit the grey-body spectrum, or (b) the 250 μm measurement only. Blue squares correspond to objects detected in the radio above $2\sigma$, and red circles are those below this detection threshold. Unfilled symbols correspond to the FIR data being below $2\sigma$. The dashed lines of best-fit are determined using all measured values, and are given by (a) SFR($L_{\rm FIR}$) = $(10.01 \pm 12.16) \log_{10}[M_{\rm BH}] - (7.44 \pm 109.98)$ and (b) SFR($L_{\rm FIR}$) = $(-27.34 \pm 16.39) \log_{10}[M_{\rm BH}] + (262.79 \pm 146.87)$. The associated coefficient of determination is shown in the bottom left-hand corner of each plot. (Note that for these linear regression analyses, we approximate the uncertainties in the measurements – which are non-Gaussian – by using the average of the upper and lower 1-$\sigma$ uncertainty in SFR($L_{\rm FIR}$).) For two RQQs, the 250 μm measurement is negative, leading to a poor determination of $L_{\rm FIR}$ via single-band fitting in the FIR. As a result, the uncertainty in SFR($L_{\rm FIR}$) for these objects strongly influences the line of best-fit for the single-band case. For interest, we repeat the linear regression analyses with these two RQQs removed, resulting in the dotted lines of best-fit, (a) SFR($L_{\rm FIR}$) = $(8.15 \pm 12.49) \log_{10}[M_{\rm BH}] + (9.99 \pm 113.06)$ and (b) SFR($L_{\rm FIR}$) = $(3.32 \pm 15.55) \log_{10}[M_{\rm BH}] + (40.71 \pm 136.75)$. Arrows represent 2-$\sigma$ upper limits in the SFR, for quasars having a 250 μm flux-density $< 2\sigma$. Overplotted are the median SFRs (black diamonds), derived using all objects, binned in $M_{\rm BH}$. Horizontal error-bars indicate the ranges of the $M_{\rm BH}$ bins, and uncertainties on the median SFRs are given by vertical error-bars.

# 8 RELATIONSHIPS BETWEEN RADIO AND OPTICAL LUMINOSITIES

In this section we investigate how star formation and black-hole accretion, quantified by their contribution to the total radio luminosity, correlate with optical luminosity. Firstly, we present the accretion-related radio luminosity ($L_{1.5\,\text{GHz, acc}}$) against the star formation-related radio luminosity ($L_{1.5\,\text{GHz, SF}}$) in Fig. 10. A linear regression analysis gives the coefficient of determination, $R^2 = 0.08$, which indicates that there is little correlation between accretion and star formation. In contrast, the Kendall statistic ($\tau = 0.37$ with $p$-value = $6.1 \times 10^{-3}$) indicates that there is statistically-significant evidence for a correlation. However, due to the different combinations of whether or not an object is detected in the radio and/or the FIR, the data include a mixture of 2-$\sigma$ lower and upper limits in $L_{1.5\,\text{GHz, acc}}$, for the non-detections. Since the *bhkmethod* task cannot accommodate a mixture of limits in a single variable, we needed to fix 9 of the RQQs at their lower limit for this Kendall test, and treat these 9 values as though they are detections. Doing so renders the test unreliable, but we still provide it for completeness. (Note that 9 is the minimum number of limits that need to be fixed in order for this Kendall test to use only detected values and one type of limit.) As before, we repeat the test without any limit values, this time using the subset of RQQs that are detected in both the radio and the FIR. The result is a Kendall statistic of $\tau = 0.28$, having no statistical significance ($p$-value = $3.2 \times 10^{-1}$, Table 2). Note that the larger $p$-value for this test may in part be due to the smaller number of objects used (i.e. 26 RQQs instead of the full sample of 70 RQQs). In conclusion, these analyses do not provide sufficient reliable information about how accretion and star formation may be connected via feedback mechanisms (if at all).

The radio emission due to star formation is shown against the absolute $i$-band magnitude, $M_i$, in the upper panel of Fig. 11. Linear regression analysis indicates (via $R^2 = 0.07$) that the trend is marginally stronger than that seen between the *total* radio luminosity and $M_i$ (Fig. 5). The reason could be that the greater scatter introduced by the accretion-related radio emission, as evident in the lower panel of Fig. 11, has been removed. However, to test the correlation robustly, we perform a Kendall test between $L_{1.5\,\text{GHz, SF}}$ and $M_i$. This gives $\tau = -0.47$ and $p$-value = $2.8 \times 10^{-3}$, indicating a significant anti-correlation. (For reference, repeating the test using only FIR-detected RQQs, and so excluding the objects with upper limits in $L_{1.5\,\text{GHz, SF}}$, results in $\tau = -0.33$ and $p$-value = $9.5 \times 10^{-2}$.) Similarly, we carry out linear regression analysis and Kendall tests for the variation in $L_{1.5\,\text{GHz, acc}}$ with $M_i$. Whilst the coefficient of determination is only $R^2 = 0.03$, there is significant evidence of anti-correlation between the two properties ($\tau = -0.59$, $p$-value $< 1.0 \times 10^{-4}$, for the full sample). That is, a greater amount of accretion-connected radio emission is associated with quasars having a brighter optical luminosity. This is expected, as the quasar's optical luminosity acts as a proxy for the accretion rate. However, we caution that this Kendall test requires the use of 9 fixed limits (to avoid a mixture of upper and lower limits in $L_{1.5\,\text{GHz, acc}}$), and so is not reliable. When only the radio-detected, FIR-detected RQQs are considered for the Kendall test, the anti-correlation is no longer evident at a statistically-significant level ($\tau = -0.63$, $p$-value = $2.3 \times 10^{-2}$, Table 2). This may be due to the smaller number of objects used for the test, but also influenced by this subset of RQQs occupying a more-restricted range in $M_i$.

Next we calculate $L_{1.5\,\text{GHz, SF}}/L_{1.5\,\text{GHz}}$, and study how this varies with $M_i$ (Fig. 12). This ratio is the fraction of the total radio luminosity that is related to star formation, and is unconstrained for objects that are undetected in *both* the radio and the FIR (red, unfilled circles in Fig. 12). For these quasars, the error-bars are used to indicate the possible range of (physical) values, from the lower limit of 0.0 (i.e. none of the radio emission is due to star formation) to the upper limit of 1.0 (i.e. all of the radio emission is due to star formation). Note, however, that there is one object with $L_{1.5\,\text{GHz, SF}}/L_{1.5\,\text{GHz}} > 1.0$, which corresponds to the single quasar lying to the left of the FIRC in Fig. 7.

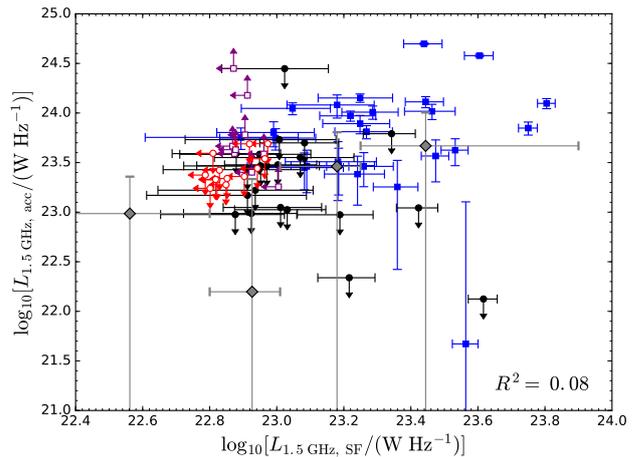

**Figure 10.** The accretion luminosity, $L_{1.5\,\text{GHz, acc}}$, against the star-formation luminosity, $L_{1.5\,\text{GHz, SF}}$. Squares (blue and purple) correspond to objects detected in the radio above 2$\sigma$, and circles (red and black) are those below this detection threshold. Unfilled symbols correspond to the FIR data being below 2$\sigma$. Arrows represent 2-$\sigma$ upper (and lower) limits in $L_{1.5\,\text{GHz, acc}}$ or $L_{1.5\,\text{GHz, SF}}$, for quasars undetected at the 2-$\sigma$ level in the JVLA images and/or the 250 μm photometry, respectively. A linear regression analysis, between $L_{1.5\,\text{GHz, acc}}$ and $L_{1.5\,\text{GHz, SF}}$, gives the coefficient of determination shown in the bottom right-hand corner. (Note that this analysis uses just the derived values and not the uncertainties in $L_{1.5\,\text{GHz, acc}}$ and $L_{1.5\,\text{GHz, SF}}$, due to them being highly correlated with one another.) Overplotted are the median luminosities (grey diamonds), derived using all objects, binned in $L_{1.5\,\text{GHz, SF}}$. Horizontal error-bars indicate the ranges of the $L_{1.5\,\text{GHz, SF}}$ bins, and uncertainties on the median accretion luminosities are given by the vertical error-bars.

Little trend can be seen in $L_{1.5\,\text{GHz, SF}}/L_{1.5\,\text{GHz}}$ with optical luminosity (Fig. 12), and we investigate this further via two Kendall correlation tests. Taking a conservative approach, the first test uses the upper limit ($L_{1.5\,\text{GHz, SF}}/L_{1.5\,\text{GHz}} = 1.0$) for the objects with unconstrained fractions. Given that these objects tend to be distributed towards the faint end of the $M_i$ range, it is unsurprising that the Kendall test results in a positive correlation ($\tau = 0.28$). However, this is not statistically significant ($p$-value = $1.8 \times 10^{-2}$) and is likely affected by 22 of the RQQs having their limit values treated as detected values. (As before, this is the minimum number of fixed limits required to avoid a mixture of upper and lower limits in a single variable – in this case, $L_{1.5\,\text{GHz, SF}}/L_{1.5\,\text{GHz}}$. Note that none of the fixed limits are associated with RQQs that have unconstrained fractions, meaning that the upper limits for these RQQs are still treated as upper limits for the test.)

For comparison, the second Kendall test uses the lower limits ($L_{1.5\,\text{GHz, SF}}/L_{1.5\,\text{GHz}} = 0.0$), resulting in $\tau = 0.32$ and $p$-value = $2.5 \times 10^{-3}$ (Table 2). This time the test requires just 9 limits to be fixed, and (again) none of these are associated with RQQs that have unconstrained fractions (i.e. their lower limits are still treated as lower limits for the test). Since the lower and upper limits for the unconstrained fractions allow us to explore the extremities of the correlation, between $L_{1.5\,\mathrm{GHz,SF}}/L_{1.5\,\mathrm{GHz}}$ and $M_i$, the 'true' $\tau$ value is expected to lie between 0.28 and 0.32 (although we do not interpret the former value as significant, and caution that a greater number of fixed limits were used in its determination). In addition, we perform a Kendall test using only the RQQs detected in both the radio and the FIR. Although the result is not statistically significant ($p$-value = $2.3 \times 10^{-2}$), the Kendall statistic of $\tau = 0.63$ confirms a positive correlation between $L_{1.5\,\mathrm{GHz,SF}}/L_{1.5\,\mathrm{GHz}}$ and $M_i$. Since a smaller value of $L_{1.5\,\mathrm{GHz,SF}}/L_{1.5\,\mathrm{GHz}}$ corresponds to a larger fraction of the radio emission being related to the AGN, these results hint towards the 'accretion fraction' ($L_{1.5\,\mathrm{GHz,acc}}/L_{1.5\,\mathrm{GHz}}$) increasing with higher optical luminosity.

Furthermore, it is interesting that the brightest bin in $M_i$ is the one that contains the greatest proportion of RQQs with a 2-$\sigma$ detection in the radio. In order for the quasars to be this optically bright, they may be accreting gas at a very high rate, since this leads to greater thermal emission from the accretion disc, which dominates the $i$ band. Alternatively, these RQQs may have a higher radiative efficiency ($\epsilon$), leading to a greater luminosity ($L$) for a given accretion rate ($\dot{M}$): $L = \epsilon \dot{M} c^2$, where $c$ = the speed of light. It is therefore reasonable to suggest that the median ratio of $L_{1.5\,\mathrm{GHz,SF}}/L_{1.5\,\mathrm{GHz}}$ in this bin is lower than the others because some mechanism is resulting in higher accretion rates or accretion that is more radiatively-efficient. The radio emission related to accretion ($L_{1.5\,\mathrm{GHz,acc}}$) is therefore a larger fraction of the total radio luminosity, compared to the fraction for other bins in $M_i$.

## 9 DISCUSSION

In this paper we have established that, for the majority of RQQs in our sample, the dominant source of radio emission is connected to the AGN rather than star formation in the host galaxy. We have also found evidence that, in radio-detected, FIR-detected RQQs, there is not a significant correlation between the radio emission due to star formation and the radio emission due to accretion (Fig. 10, with $\tau = 0.28$ and $p$-value = $3.2 \times 10^{-1}$). The reason for this finding could be due to the different timescales associated with AGN activity and star formation, leading to a delay between the two (Wild et al. 2010) that may be particularly pronounced for radio-quiet objects.

Considering this further, we now discuss the possible origins of the accretion-connected radio emission in RQQs. A Kendall test using the full sample, albeit with 9 lower limits treated as detected values, shows that there is a significant correlation between the accretion-connected radio-luminosity and the optical luminosity (Table 2). This is unsurprising, given that it may be expected for both to be closely associated with the accretion disc. Meanwhile, a weaker (though still statistically-significant) correlation exists between $L_{1.5\,\mathrm{GHz,SF}}$ and optical luminosity. [Note that a stronger correlation is exhibited in radio-loud quasars (Kalfountzou et al. 2012, 2014), possibly brought about by jet-induced star-formation (e.g. Croft et al. 2006).] As a common gas-reservoir is thought to fuel both star formation and accretion, different efficiencies of these two processes may explain the differing strengths in the correlation with $M_i$. Furthermore, with $M_i$ acting as a proxy for accretion rate, other factors must be at work to explain the scatter in $L_{1.5\,\mathrm{GHz,acc}}$ (lower panel of Fig. 11).

Blandford & Payne (1982) and Reynolds et al. (2006) argue that the power of radio jets (in radio-loud quasars and X-ray binaries, respectively) scales with accretion rate, given that the accretion of magnetic flux onto the black hole may factor into whether or not a radio jet is produced. Although we may expect the observed trend of $L_{1.5\,\mathrm{GHz,acc}}$ with accretion rate to be more pronounced, we cannot rule out the presence of jets in RQQs (e.g. Blundell & Rawlings 2001) as they may be too small to be resolved in the JVLA images (e.g. Kellermann et al. 2004). Regarding the scatter in the lower panel of Fig. 11, magnetic fields may play a role, leading to jets that boost the value of $L_{1.5\,\mathrm{GHz,acc}}$ in some objects.

Furthermore, the environmental density will dictate how much radio emission results from such jets (if present). The reason is that denser material will interact more strongly with the jet, thus rapidly decelerating electrons and producing more radio emission. This is substantiated by the hydrodynamical numerical modelling of radio-galaxy lobes by Hardcastle & Krause (2013). However, they focus on FRII objects, and simply scaling down the interaction between jets and the environment may be inappropriate for RQQs. Along these lines, Yee & Green (1984) and Falder et al. (2010) find that RQQs occupy less-dense environments compared to their radio-loud counterparts.

In addition, the magnetic field in or surrounding the accretion disc may not be of sufficient strength to produce high jet-power. This could point towards a different mechanism to that in radio-loud quasars, producing significant radio emission. For example, it is possible that radio emission from shock fronts, associated with quasar outflows, is making a contribution. Zakamska & Greene (2014) infer that the outflows are driven by the radiative output of the quasar, and so the resulting radio luminosity should scale with the optical luminosity. Assuming that the latter is an adequate tracer of the AGN's bolometric luminosity, such a scaling is used by Nims et al. (2015) to predict the amount of synchrotron emission due to outflows, which they find to be in agreement with typical radio emission from RQQs.

The AGN-related radio emission could instead be dominated by disc winds from the outermost regions of the accretion disc (as observed in microquasars, e.g. Blundell et al. 2001). These winds may arise when photo-ionised plasma, irradiated by the central region of the X-ray binary or quasar, has a thermal velocity that is greater than the local escape velocity (Proga & Kallman 2002). For velocities below this threshold, the plasma remains bound to the accretion disc. Clearly, this implies that the strength of the disc wind depends on the bolometric luminosity, which traces the central black-hole's accretion rate (Shakura & Sunyaev 1973; Kuncic & Bicknell 2004, 2007). Therefore, disc winds (like unresolved jets) may explain our observation of higher values of $L_{1.5\,\mathrm{GHz,acc}}$ with the brightest optical luminosities. In fact, disc winds associated with high accretion-rates appear to suppress the jet-production mechanism in X-ray binaries (e.g. Neilsen & Lee 2009; Ponti et al. 2012). Further support for this scenario is provided by King et al. (2013), who find that there is a transition from jet power to wind power at a particular Eddington ratio. We also note that, if applicable for RQQs, jet suppression could increase the effectiveness of negative feedback from disc winds on star formation (Garofalo et al. 2016).

Another suggestion, by Laor & Behar (2008), is that synchrotron emission in RQQs could be due to magnetic reconnection events accelerating electrons in the corona above the disc. This is developed further by Raginski & Laor (2016), who argue that the spectral index and variability of RQQs favours a coronal origin over an AGN-driven wind. Further investigation of the above explanations is beyond the scope of this work, but a combination of these different processes may be the reason for the large variation seen in $L_{1.5\,\mathrm{GHz,acc}}$ as a function of $M_i$.

Finally, we note that the faintest $M_i$ bin in Fig. 11 contains numerous objects with negative radio flux-densities, and so the median $L_{1.5\,\text{GHz, acc}}$ value is itself negative. With deeper radio data the detection rate should improve, allowing for better investigation of trends between the radio and the optical at the faint end.

## 10 CONCLUSIONS

Using a sample selected over a single epoch ($0.9 < z < 1.1$) and spanning two orders of magnitude in optical luminosity, we have investigated the star-formation and accretion properties of RQQs in the *Spitzer-Herschel* Active Galaxy Survey. The narrow redshift range allows any evolutionary effects to be decoupled from our findings, breaking the degeneracy between luminosity and redshift that is inherent in most samples. Previous studies that used the same sample needed to invoke stacking analyses, as the radio data were not of sufficient sensitivity. The current work uses new JVLA images to uncover the following results:

(i) 35 of the 70 objects are detected at a 2-$\sigma$ level in the 1.5 GHz radio images, with the median flux-density for the whole sample determined to be $85.0 \pm 17.2\,\mu\text{Jy beam}^{-1}$.

(ii) A linear regression analysis between the radio luminosity and $M_i$ indicates that correlation between the two is poor, although a weak trend can be seen by eye, with the optically-bright quasars having the highest radio luminosities. This is confirmed by a Kendall rank correlation test, indicating a significant anti-correlation.

(iii) Far-infrared luminosities are used to estimate individual star-formation rates (SFRs) for the quasars, which are then compared against SFRs derived from the radio luminosity. The latter are based on the assumption that all of the radio emission is a consequence of star-formation processes, but we show that this is not the case.

(iv) The inclusion of black-hole masses in the analysis indicates that quasars harbouring more-massive black holes do not necessarily have higher SFRs. This may be expected from more-massive black holes residing in galaxies with larger stellar mass, which in turn are more likely to host larger reservoirs of gas. One explanation for the lack of correlation is the role of AGN feedback, with star formation being suppressed as a result of accretion-related heating of the gas. Another is simply that black-hole mass and SFR are not related.

(v) The monochromatic luminosity at rest-frame $125\,\mu\text{m}$, $L_{125\,\mu\text{m}}$, is a reliable tracer of the far-infrared emission for this sample. It is therefore used to describe the far-infrared to radio correlation (FIRC), via the dimensionless parameter $q_{125}$. As found by White et al. (2015), emission from star formation alone cannot explain the total radio emission for the majority of RQQs, suggesting that accretion still makes an important contribution.

(vi) Exploiting the FIRC, we divide the total radio emission for each source ($L_{1.5\,\text{GHz}}$) into its constituent components: that related to star formation, $L_{1.5\,\text{GHz, SF}}$, and that related to accretion, $L_{1.5\,\text{GHz, acc}}$. The latter is used to quantify the contribution of the AGN to the total radio luminosity. Given a 2-$\sigma$ detection level in both the radio and the FIR, 92 per cent of the RQQs are AGN-dominated (having $L_{1.5\,\text{GHz, acc}} > 0.5\,L_{1.5\,\text{GHz}}$), and the accretion process accounts for 80 per cent of the radio luminosity when summed across the objects. This proportion of the radio luminosity increases to 92 per cent when considering the upper limits in $L_{1.5\,\text{GHz, acc}}$, for objects lying over 2$\sigma$ away from the FIRC (in the direction of more radio-luminous sources, see Fig. 7).

(vii) The radio emission connected with star formation appears to be anti-correlated with $M_i$, as suggested by the Kendall statistic. Meanwhile, $L_{1.5\,\text{GHz, acc}}$ shows large scatter when plotted against $M_i$, but an anti-correlation between the two properties is also shown to be statistically significant when upper limits are considered (although lower limits need to be treated as detected values for this Kendall test). However, the degree of correlation between optical luminosity and the fraction of radio emission due to star formation ($L_{1.5\,\text{GHz, SF}}/L_{1.5\,\text{GHz}}$) cannot be determined explicitly, due to objects with unconstrained fractions dominating the faint end of the range in $M_i$. Deeper radio and FIR data, a larger sample size, and additional multi-wavelength data may help to disentangle the various feedback mechanisms at work, if they are indeed playing a significant role.

The new generation of radio telescopes, due to come online during the next decade, will lead to a number of direct detections of radio-quiet AGN that greatly exceeds existing samples, such as that used for the current work. As a result it will be possible to investigate accretion-related radio emission and star formation-related radio emission further, including their trend with both redshift and optical luminosity. Doing so will provide additional insight into how these processes influence the way in which galaxies evolve, and how the mechanisms behind the production of radio emission in RQQs may differ from those present in their radio-loud counterparts.

## ACKNOWLEDGEMENTS

SVW would like to thank Ian Heywood, Juergen Ott and Kim McAlpine for their advice whilst carrying out reduction and imaging of the JVLA data. In addition, we thank staff at the NRAO Helpdesk, who helped to resolve the running of the JVLA pipeline over problematic scheduling blocks. We also thank the referee, David Rosario, whose suggestions improved the robustness of the statistical analysis. SVW acknowledges support provided through a UK Science and Technology Facilities Council (STFC) studentship, MJH acknowledges support from the UK STFC [ST/M001008/1], and AV acknowledges support from the Leverhulme Trust in the form of a Research Fellowship.

The National Radio Astronomy Observatory is a facility of the National Science Foundation operated under cooperative agreement by Associated Universities, Inc.

The *Herschel* spacecraft was designed, built, tested, and launched under a contract to ESA managed by the *Herschel*/*Planck* Project team by an industrial consortium under the overall responsibility of the prime contractor Thales Alenia Space (Cannes), and including Astrium (Friedrichshafen) responsible for the payload module and for system testing at spacecraft level, Thales Alenia Space (Turin) responsible for the service module, and Astrium (Toulouse) responsible for the telescope, with in excess of a hundred subcontractors.

HIPE is a joint development by the *Herschel* Science Ground Segment Consortium, consisting of ESA, the NASA *Herschel* Science Center, and the HIFI, PACS and SPIRE consortia.

Funding for the Sloan Digital Sky Survey IV has been provided by the Alfred P. Sloan Foundation, the U.S. Department of Energy Office of Science, and the Participating Institutions. SDSS-IV acknowledges support and resources from the Center for High-Performance Computing at the University of Utah. The SDSS web site is www.sdss.org.

# APPENDIX A: RADIO MEASUREMENTS FOR THE RQQS

Presented over two pages is a table of radio measurements, obtained from newly-reduced JVLA data, for the 70 RQQs in the *Spitzer-Herschel* Active Galaxy Survey (Table A1).

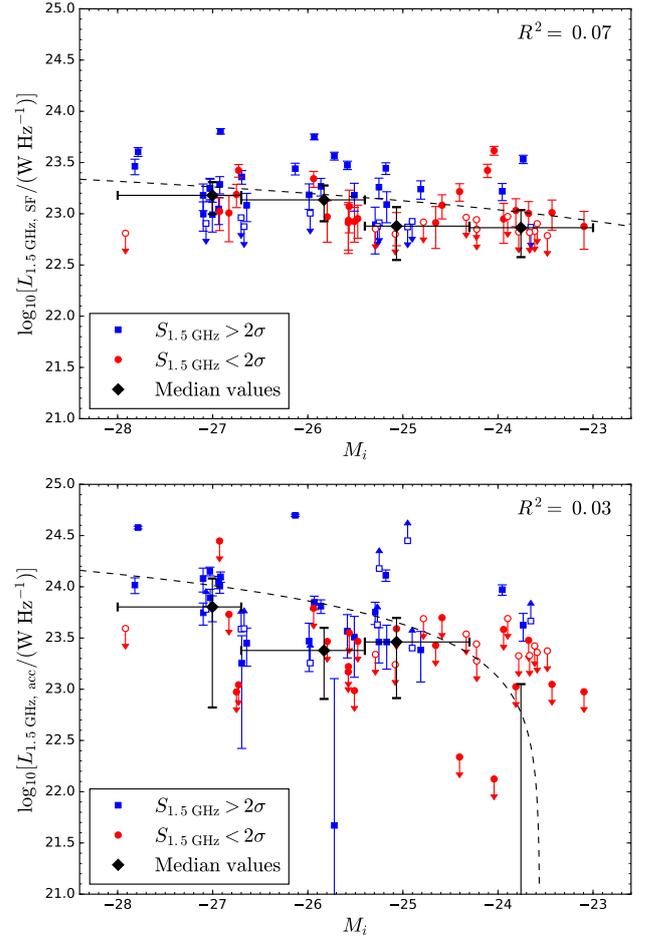

**Figure 11.** The star-formation luminosity, $L_{1.5\,\mathrm{GHz,\,SF}}$ (upper panel), and the accretion luminosity, $L_{1.5\,\mathrm{GHz,\,acc}}$ (lower panel), against the absolute $i$-band magnitude, $M_i$. Blue squares correspond to objects detected in the radio above $2\sigma$, and red circles are those below this detection threshold. Unfilled symbols correspond to the FIR data being below $2\sigma$. Arrows in the upper plot represent $2$-$\sigma$ upper limits in $L_{1.5\,\mathrm{GHz,\,SF}}$, for quasars having a flux density measured at $250\,\mu\mathrm{m} < 2\,\sigma$. In the lower plot, the arrows indicate whether the value of $L_{1.5\,\mathrm{GHz,\,acc}}$ is either an upper or lower limit (again at $2\sigma$), dependent on whether the object is undetected in both the radio and the FIR, or undetected in the FIR alone. The lines of best-fit are given by $L_{1.5\,\mathrm{GHz,\,SF}} = (-2.45 \pm 1.14) \times 10^{22} M_i - (4.78 \pm 2.89) \times 10^{23}$ (upper panel) and $L_{1.5\,\mathrm{GHz,\,acc}} = (-2.99 \pm 0.84) \times 10^{23} M_i - (7.05 \pm 2.11) \times 10^{24}$ (lower panel), and the associated coefficient of determination is shown in the top right-hand corner of each panel. (Uncertainties in $L_{1.5\,\mathrm{GHz,\,SF}}$ and $L_{1.5\,\mathrm{GHz,\,acc}}$ are used for these fits.) The dashed lines are the result of converting these best-fit lines into log–linear space. Overplotted are the median luminosities (black diamonds), derived using all objects, binned in $M_i$. The horizontal error-bars indicate the ranges of the $M_i$ bins (Table 1), and uncertainties on the median radio-luminosities are given by the vertical error-bars. Note that the values of the luminosities, even if negative, are used for the linear regression analysis and the calculation of the median luminosities, rather than the limit values.

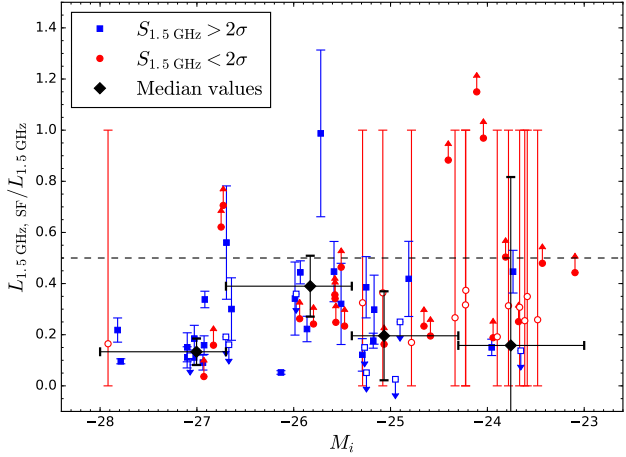

**Figure 12.** The fraction of the radio luminosity that is related to star formation, $L_{1.5\,\text{GHz, SF}}/L_{1.5\,\text{GHz}}$, as a function of the absolute *i*-band magnitude, $M_i$. Blue squares correspond to objects detected in the radio above $2\sigma$, and red circles are those below this detection threshold. Unfilled symbols correspond to the FIR data being below $2\sigma$. The arrows represent 2-$\sigma$ upper (and lower) limits in the value of $L_{1.5\,\text{GHz, SF}}/L_{1.5\,\text{GHz}}$, and objects that are undetected in both the radio and the FIR have error-bars spanning the full (physical) range from 0.0 to 1.0 (i.e. the fraction is unconstrained). The dashed line is to guide the eye where $L_{1.5\,\text{GHz, SF}} = L_{1.5\,\text{GHz, acc}}$. Overplotted are the median ratios (black diamonds), derived using all objects, binned in $M_i$. The horizontal error-bars indicate the ranges of the $M_i$ bins (Table 1), and uncertainties on the median ratios are given by the vertical error-bars.

**Table A1.** Radio measurements, taken from JVLA imaging, for the sample of 70 radio-quiet quasars. The theoretical noise is calculated according to the standard sensitivity-equation. 'Measured rms' refers to the noise over a defined region, which excludes sources in the field, and SNR = peak flux-density / measured rms.

| Target's full name | R.A. (hms) | Dec. (dms) | Theoretical rms (μJy beam$^{-1}$) | Measured rms (μJy beam$^{-1}$) | Peak flux-density (μJy beam$^{-1}$) | SNR |
|---|---|---|---|---|---|---|
| SDSS003146.07+134629.6 | 00:31:46.07 | +13:46:30.0 | 14.8 | 21.9 | -13.0 | -0.6 |
| SDSS023540.90+001038.9 | 02:35:40.90 | +00:10:39.2 | 15.6 | 29.4 | 143.2 | 4.9 |
| SDSS073802.37+383116.3 | 07:38:02.35 | +38:31:16.5 | 20.0 | 31.3 | 422.8 | 13.5 |
| SDSS074729.24+434607.5 | 07:47:29.16 | +43:46:07.8 | 22.1 | 26.9 | 5.0 | 0.2 |
| SDSS075058.21+421617.0 | 07:50:58.20 | +42:16:16.9 | 19.8 | 29.0 | -8.3 | -0.3 |
| SDSS075222.91+273823.2 | 07:52:22.89 | +27:38:23.0 | 19.7 | 38.2 | 141.4 | 3.7 |
| SDSS075339.84+250137.9 | 07:53:39.81 | +25:01:37.8 | 19.4 | 46.7 | 56.5 | 1.2 |
| SDSS082229.78+442705.2 | 08:22:29.76 | +44:27:05.2 | 19.5 | 40.5 | 99.1 | 2.4 |
| SDSS083115.89+423316.6 | 08:31:15.86 | +42:33:16.5 | 19.2 | 45.7 | 185.5 | 4.1 |
| SDSS084723.67+011010.4 | 08:47:23.64 | +01:10:10.3 | 15.1 | 46.5 | 42.9 | 0.9 |
| SDSS090153.42+065915.6 | 09:01:53.42 | +06:59:15.3 | 14.8 | 25.6 | 313.3 | 12.3 |
| SDSS091216.88+420314.2 | 09:12:16.87 | +42:03:14.3 | 21.2 | 48.6 | -43.2 | -0.9 |
| SDSS092257.86+444651.8 | 09:22:57.86 | +44:46:52.1 | 20.5 | 83.2 | 136.2 | 1.6 |
| SDSS092753.52+053637.0 | 09:27:53.52 | +05:36:36.8 | 14.6 | 28.7 | 57.7 | 2.0 |
| SDSS092829.86+504836.6 | 09:28:29.78 | +50:48:36.3 | 22.4 | 31.1 | 346.2 | 11.1 |
| SDSS093023.28+403111.0 | 09:30:23.28 | +40:31:11.1 | 22.1 | 35.4 | 90.3 | 2.6 |
| SDSS093303.50+460440.2 | 09:33:03.48 | +46:04:39.8 | 19.5 | 27.0 | 86.5 | 3.2 |
| SDSS093759.44+542427.3 | 09:37:59.35 | +54:24:27.2 | 22.6 | 43.3 | 68.0 | 1.6 |
| SDSS094811.89+551726.5 | 09:48:11.85 | +55:17:26.4 | 24.8 | 51.5 | -10.7 | -0.2 |
| SDSS100730.47+050942.3 | 10:07:30.48 | +05:09:42.0 | 17.1 | 35.4 | 218.0 | 6.2 |
| SDSS100835.81+513927.8 | 10:08:35.83 | +51:39:27.9 | 16.9 | 22.4 | 36.4 | 1.6 |
| SDSS100906.35+023555.3 | 10:09:06.33 | +02:35:55.4 | 17.4 | 59.2 | 256.3 | 4.3 |
| SDSS102005.99+033308.5 | 10:20:05.97 | +03:33:08.4 | 19.9 | 48.7 | 201.7 | 4.1 |
| SDSS102111.57+611415.0 | 10:21:11.56 | +61:14:15.2 | 21.5 | 32.9 | 40.6 | 1.2 |
| SDSS102349.40+522151.2 | 10:23:49.39 | +52:21:51.2 | 20.2 | 39.1 | 107.1 | 2.7 |
| SDSS103347.32+094039.0 | 10:33:47.30 | +09:40:39.0 | 17.5 | 33.4 | 268.1 | 8.0 |
| SDSS103525.05+580335.6 | 10:35:25.05 | +58:03:35.6 | 17.3 | 79.6 | 158.5 | 2.0 |
| SDSS103829.74+585204.1 | 10:38:29.73 | +58:52:04.1 | 17.6 | 28.8 | 303.6 | 10.5 |
| SDSS103855.33+575814.7 | 10:38:55.32 | +57:58:14.7 | 17.3 | 60.8 | 84.4 | 1.4 |
| SDSS104114.18+590219.4 | 10:41:14.18 | +59:02:19.5 | 20.2 | 46.8 | 3.1 | 0.1 |
| SDSS104239.66+583231.0 | 10:42:39.64 | +58:32:31.0 | 20.0 | 41.4 | -50.4 | -1.2 |
| SDSS104355.47+593054.0 | 10:43:55.46 | +59:30:54.0 | 16.7 | 25.0 | 21.1 | 0.8 |
| SDSS104537.69+484914.6 | 10:45:37.68 | +48:49:14.6 | 21.8 | 54.2 | -27.8 | -0.5 |
| SDSS104659.37+573055.6 | 10:46:59.37 | +57:30:55.8 | 16.8 | 26.9 | 74.7 | 2.8 |
| SDSS104859.67+565648.6 | 10:48:59.66 | +56:56:48.6 | 21.3 | 30.5 | 147.4 | 4.8 |

**Table A1.** *Continued* – Radio measurements, taken from JVLA imaging, for the sample of 70 radio-quiet quasars. The theoretical noise is calculated according to the standard sensitivity-equation. 'Measured rms' refers to the noise over a defined region, which excludes sources in the field, and SNR = peak flux-density / measured rms.

| Target's full name | R.A. (hms) | Dec. (dms) | Theoretical rms ($\mu$Jy beam$^{-1}$) | Measured rms ($\mu$Jy beam$^{-1}$) | Peak flux-density ($\mu$Jy beam$^{-1}$) | SNR |
|---|---|---|---|---|---|---|
| SDSSJ104930.46+592032.6 | 10:49:30.45 | +59:20:32.7 | 17.4 | 26.9 | 663.5 | 24.7 |
| SDSSJ104935.76+554950.6 | 10:49:35.76 | +55:49:50.5 | 22.6 | 40.1 | -8.6 | -0.2 |
| SDSSJ105408.88+042650.4 | 10:54:08.88 | +04:26:50.4 | 17.9 | 36.6 | 29.8 | 0.8 |
| SDSSJ112317.52+051804.0 | 11:23:17.49 | +05:18:03.9 | 15.6 | 342.0 | -875.6 | -2.6 |
| SDSSJ115027.25+665848.0 | 11:50:27.21 | +66:58:48.1 | 18.9 | 28.7 | 143.0 | 5.0 |
| SDSSJ122832.94+603735.1 | 12:28:32.92 | +60:37:35.1 | 18.0 | 28.5 | 26.5 | 0.9 |
| SDSSJ123059.71+101624.8 | 12:30:59.71 | +10:16:24.5 | 15.5 | 26.1 | 85.6 | 3.3 |
| SDSSJ125659.93+042734.4 | 12:56:59.90 | +04:27:34.6 | 15.2 | 24.1 | 931.8 | 38.7 |
| SDSSJ132957.15+540505.9 | 13:29:57.14 | +54:05:06.0 | 18.4 | 39.7 | 326.0 | 8.2 |
| SDSSJ133713.06+610749.0 | 13:37:13.00 | +61:07:49.0 | 18.1 | 30.1 | 36.8 | 1.2 |
| SDSSJ133733.30+590622.6 | 13:37:33.28 | +59:06:22.8 | 18.4 | 30.9 | 186.0 | 6.0 |
| SDSSJ135823.99+021343.8 | 13:58:23.97 | +02:13:44.0 | 14.7 | 51.2 | 3.5 | 0.1 |
| SDSSJ142124.65+423003.2 | 14:21:24.67 | +42:30:03.1 | 14.1 | 21.5 | 196.0 | 9.1 |
| SDSSJ142817.30+502712.6 | 14:28:17.30 | +50:27:12.7 | 14.9 | 33.8 | 106.5 | 3.1 |
| SDSSJ145503.47+014209.0 | 14:55:03.45 | +01:42:09.2 | 16.0 | 24.1 | 35.6 | 1.5 |
| SDSSJ145506.12+562935.6 | 14:55:06.09 | +56:29:35.6 | 14.6 | 23.2 | 86.9 | 3.7 |
| SDSSJ151520.56+004739.3 | 15:15:20.54 | +00:47:39.4 | 19.0 | 42.7 | 416.8 | 9.8 |
| SDSSJ151921.85+535842.3 | 15:19:21.84 | +53:58:42.2 | 13.9 | 25.4 | 24.6 | 1.0 |
| SDSSJ155436.25+320408.4 | 15:54:36.24 | +32:04:08.5 | 14.6 | 25.7 | 14.5 | 0.6 |
| SDSSJ155650.41+394542.8 | 15:56:50.40 | +39:45:42.8 | 14.7 | 32.0 | 100.6 | 3.1 |
| SDSSJ163225.56+411852.0 | 16:32:25.56 | +41:18:52.4 | 18.6 | 25.6 | 47.1 | 1.8 |
| SDSSJ163306.12+401747.0 | 16:33:06.12 | +40:17:47.5 | 17.7 | 26.7 | 11.6 | 0.4 |
| SDSSJ163408.64+331242.1 | 16:34:08.64 | +33:12:42.1 | 15.3 | 27.6 | 116.4 | 4.2 |
| SDSSJ163930.82+410013.2 | 16:39:30.81 | +41:00:13.6 | 17.2 | 26.5 | 20.3 | 0.8 |
| SDSSJ164617.17+364509.4 | 16:46:17.16 | +36:45:09.6 | 17.1 | 27.9 | -9.0 | -0.3 |
| SDSSJ165231.30+353615.9 | 16:52:31.29 | +35:36:15.9 | 17.9 | 26.1 | 30.1 | 1.2 |
| SDSSJ171005.53+644843.0 | 17:10:05.52 | +64:48:42.9 | 14.6 | 23.0 | 293.2 | 12.7 |
| SDSSJ171145.53+601318.4 | 17:11:45.52 | +60:13:18.6 | 14.1 | 27.0 | 30.3 | 1.1 |
| SDSSJ171330.21+644253.0 | 17:13:30.24 | +64:42:53.0 | 17.8 | 26.0 | 1104.9 | 42.4 |
| SDSSJ171704.69+281400.6 | 17:17:04.68 | +28:14:00.7 | 16.5 | 34.7 | 263.5 | 7.6 |
| SDSSJ171732.94+594747.7 | 17:17:32.92 | +59:47:47.5 | 14.2 | 20.8 | 97.1 | 4.7 |
| SDSSJ172130.96+584404.1 | 17:21:30.96 | +58:44:04.7 | 14.4 | 45.1 | 45.1 | 1.0 |
| SDSSJ172310.35+595105.6 | 17:23:10.34 | +59:51:05.7 | 17.7 | 28.8 | 100.5 | 3.5 |
| SDSSJ215541.74+122818.8 | 21:55:41.73 | +12:28:18.9 | 16.2 | 65.3 | -40.7 | -0.6 |
| SDSSJ224159.43+142055.2 | 22:41:59.42 | +14:20:55.0 | 16.2 | 23.8 | 27.6 | 1.2 |